\documentclass[journal]{IEEEtran}
\usepackage{subfloat}

\usepackage[final]{graphicx}
\graphicspath{{figs}}
\usepackage{tikz}
\ifCLASSINFOpdf
\else
\fi
\usepackage[cmex10]{amsmath}
\usepackage{bm}
\usepackage{amssymb}
\usepackage{algorithm2e}
\usepackage{algorithmic}
\usepackage{xcolor}
\usepackage{subcaption}
\usepackage{cite}
\usepackage{soul}
\newcommand{\mb}{\mathbf}

\newcommand{\uvec}[1]{\hat{\mb{#1}}}

\newcommand{\rv}{\mb{r}}

\newcommand{\refp}[1]{(\ref{#1})}

\newcommand{\brcs}[1]{\lbrace #1 \rbrace}

\begin{document}
\title{The Generalized Method of Moments for Electromagnetic Boundary Integral Equations}
\author{Daniel~Dault,~\IEEEmembership{Student~Member,~IEEE,}
		  Naveen V. Nair,~\IEEEmembership{Member,~IEEE,}
		  Jie Li,~\IEEEmembership{Student Member,~IEEE,}
		  Balasubramaniam~Shanker,~\IEEEmembership{Fellow}
		  \thanks{D. Dault is with the Department of Electrical and Computer Engineering, Michigan State University, East Lansing, MI 48824-1226 email: daultdan@egr.msu.edu}
		  \thanks{N.V. Nair, J. Li and B. Shanker are with the Department of Electrical and Computer Engineering, Michigan State University}}
\maketitle

\IEEEpeerreviewmaketitle
\begin{abstract}
	The Generalized Method of Moments (GMM) is a partition of unity based technique for solving electromagnetic and acoustic boundary integral equations.  Past work on the GMM for electromagnetics was confined to geometries modeled by piecewise flat tessellations and suffered from spurious internal line charges.  In the present article, we redesign the GMM scheme and demonstrate its ability to model scattering from PEC scatterers composed of mixtures of smooth and non-smooth geometrical features.  Furthermore, we demonstrate that because the partition of unity provides function and effective geometrical continuity between patches, the GMM permits mixtures of local geometry descriptions and approximation function spaces with significantly more freedom than traditional moment methods.

\end{abstract}


\section{Introduction}

\IEEEPARstart{S}{i}gnificant effort has been exerted in recent years toward constructing discretizations of electromagnetic integral equations that model the underlying continuous problem more closely than traditional low order approaches.  The aim of these efforts is to obtain solutions with better fidelity to the continuous solution while significantly reducing the size of the descretized system and realizing commensurate savings in computational cost.  Most of the work to these ends has focused on one (or a combination) of three directions: higher order current approximating function spaces that require fewer than the traditional 10 degrees of freedom per wavelength, higher order geometry descriptions that more closely model the true continuous geometry, and mixtures of basis sets, e.g. mixing low order interpolatory basis sets with those derived from asymptotic wave representations.

Moment methods that are higher order in both current approximation spaces and geometry representation have been a subject of intense development in recent years.  Methods for higher order current representation have been traditionally based either on hierarchical higher order functions residing in Nedelec Spaces \cite{Nedelec1980,Wang1997,Graglia1997,Peterson2011}, or on variations of mapped higher order polynomial tensor products \cite{Notaros2001,Jorgensen2004,Notaros2008,Zha2012}. By reducing the number of degrees of freedom per wavelength required to discretize the integral operator, each of these methods realizes reductions in MoM system size.  However, the physical requirement that approximation functions must provide current continuity between discretization subdomains places strict limitations on the types of basis functions that may be employed, and means that the boundary integral operators generally must be discretized using a single class of basis function, e.g. only polynomials.  Some methods for mixing higher order polynomial bases with other basis classes have been developed (e.g. singularity type bases in {\cite{Graglia2008}}), but these also require that the basis set be designed to enforce current continuity.

Closely related to the issue of continuity in basis function set is that of continuity in geometry description.  Higher order surface descriptions for moment methods are generally based on smooth polynomial surface parameterizations \cite{Djordjevic2004}, which have difficulty accurately modeling geometrical singularities such as edges and tips.  Recently, representations using Non-rational B-Splines (NURBS) have been developed \cite{Liu2009,Yuan2009}.  A bottleneck with both of these methods is enforcing geometrical continuity at abutting patch edges.  As with current approximations, this generally forces the use of the same type approximating function for the parameterization of each subdomain.  

The GMM is a partition of unity method that decomposes the scatterer surface into overlapping subdomains, termed ``patches''.  As in the Generalized Finite Element Method \cite{BABUSKA1997,Duarte2000}, the partition of unity serves to decouple approximation function descriptions in neighboring subdomains.  Consequentially, GMM is able to easily incorporate higher order basis sets, higher order/mixed geometry descriptions, and arbitrary mixing of approximation function spaces.  This flexibility in approximation functions occurs because interpatch continuity is provided not by the approximating functions themselves, but by the partition of unity.  This lifting of the continuity constraint on the basis functions permits substantially more freedom in the types of approximation function spaces that may be employed; additionally, it allows mixing of different approximation spaces in adjacent subdomains over the surface of the scattering body.  Thus, in the GMM, basis sets may be freely chosen to match local current ansatz.  The partition of unity scheme also allows blending of different functional geometrical descriptions in the overlap region between subdomains, effectively providing geometrical continuity and permitting the use of distinct geometry parameterizations on neighboring subdomains.  This implies that geometry parameterizations from entirely different spaces, e.g. polynomials, conics, and flat tessellations, may be utilized in the same simulation while maintaining effective geometrical continuity. 

Although partition of unity methods have been widely employed in finite elements to address some of the challenges in basis function and geometrical continuity \cite{Oden1998,Strouboulis2000,Babuska2003,Duarte2006}, extension of these methods to integral equations has been limited, in part  because defining meaningful partitions of unity on arbitrary two dimensional manifolds residing in $\mathbb{R}^3$ is a nontrivial problem. Nonetheless, the Partition of Unity Boundary Element Method (PUBEM) \cite{PerreyDebain2003} has been developed, primarily for the solution of two dimensional acoustic Helmholtz scattering problems.  In PUBEM, traditional interpolating approximation function spaces are enriched with asymptotic plane wave bases for electrically/acoustically large simple scatterers in two dimensions \cite{Beriot2010,Peake2013}, and a sphere in three dimensions \cite{PerreyDebain2004}.  A similar approach is taken in \cite{Bruno2003a,Bruno2007a}, wherein stationary phase methods are applied to acoustics and electromagnetics problems using asymptotic representations on electrically large smooth geometries.  The approach in the present paper is significantly more general than these approaches because it applies to vectorial electromagnetics problems on arbitrarily shaped scatterers in three dimensions.  The approximation spaces utilized may be arbitrary mixtures of basis functions including, but not restricted to, the plane-wave and asymptotic type expansions employed in the PUBEM and \cite{Bruno2007a}.

Finally, we note that the Generalized Method of Moments may be technically classified as a quasi-meshless method.  Meshless methods have traditionally been confined to the finite element community, especially in the field of mechanics and mechanical engineering (e.g. \cite{Belytschko1996,Liu2010c}), although there has been increasing interest toward applying such methods to computational electromagnetics, primarily in the context of quasi-static problems \cite{Cingoski1998,Xuan2004,Bottauscio2006}, and generalized finite elements, \cite{Lu2007a,Tuncer2010}.   To the authors' knowledge, the only applications of meshless methods to high frequency electromagnetic integral equations, excepting GMM and the references in the preceding paragraph, are implementations of the traditional Moving Least Squares (MLS) method \cite{Nicomedes2009,Nicomedes2010} and the work in \cite{Tong2012}, in which both a MLS-based collocation scheme and an integral transform-based approach are given.  The Generalized Method of Moments is distinct from these methods in that it incorporates ideas from mesh-free discretization (node-based primitives, partitions of unity) to effect localization of geometry and approximation functions, but then marries these ideas with the large variety of moment method basis sets that have been developed by the electromagnetics community over the last five decades.

The present work provides a unified prescription for the algorithmic development and implementation of the Generalized Method of Moments for arbitrary PEC scatterers.  The GMM for electromagnetic integral equations was first presented in \cite{Nair2011c} for piecewise flat tessellations and has been extended to low-order (tessellated) M\"{u}ller formulations for dielectrics \cite{Nair:11} and a higher order smooth formulation for acoustics \cite{Nair2012b}. In this work, we extend electromagnetic GMM to geometries composed of mixtures of features including flat regions, tessellations, polynomial smooth patches, sharp tips, bodies of revolution, conic sections, etc.  The inclusion of local smooth geometry descriptions removes one major hurdle encountered in the piecewise flat approach in \cite{Nair2011c}, which is the appearance of spurious line charges at boundaries between non-coplanar triangles.  By introducing either smooth local geometry parameterizations or subdomain basis sets that cancel line charges by construction, the present work avoids line charges altogether.  Preliminary work on smooth local parameterizations and hybridizations with non-smooth geometrical descriptions for GMM is contained in \cite{Nair2011,Nair2011a}, and \cite{Dault2012}.

Specific contributions of this paper are:
\begin{itemize}
	\item A framework for arbitrarily mixing different classes of basis functions over the surface of a PEC scatter.
	\item A geometrical hybridization scheme wherein various functional geometry descriptions may be combined and blended in a single problem.
	\item A smooth, higher order geometry representation framework for representing arbitrary curved geometries.
	\item Hybridization with tessellations and introduction of ``sub-patch'' basis sets capable of handling geometrical singularities.
	\item Algorithms for automatically assigning approximation function types and local geometry descriptions based on local physical characteristics of the scattering geometry.
	\item Results demonstrating the flexibility of the method on several test scattering problems.  
\end{itemize}
The scattering results are designed to demonstrate mixing of geometry descriptions and basis function classes, and the corresponding reduction in system size.  To address larger problems, the method will be hybridized with the Multilevel Fast Multipole Method \cite{Song1995a} in future work.

The remainder of the paper is organized as follows.  Section \ref{s:probs} briefly outlines the problem under consideration.  The formulation of GMM for arbitrary PEC scatterers, including construction of patches starting from point clouds, the partitions of unity, local geometry parameterizations, local basis functions, and the evaluation of matrix elements, is detailed in Section \ref{s:Form}.  Results validating the method against a reference code and demonstrating application of the method to several representative geometries are presented in section \ref{s:Res}.  Finally, Section \ref{s:Con} provides some concluding remarks and future directions.

\section{Problem Statement}\label{s:probs}
The problem of interest is the computation of the scattered fields $\brcs{\mb{E}^s(\rv),\mb{H}^s(\rv)}$ due to a plane wave, characterized by the triad $\brcs{\uvec{k}^i,\mb{E}^i(\rv),\mb{H}^i(\rv)}$, impinging on a PEC object residing in free space.   The scatterer boundary $\Omega$ is equipped with a unit normal $\uvec{n}(\rv)$ defined for all $\rv \in \Omega$ except at a finite number of geometrical singularities (e.g. corners, tips, edges, etc.).  The boundary integral formulation we employ is the Combined Field Integral equation (CFIE):

\begin{equation}
	\begin{split}
	\alpha{\hat{\bf n}({\bf r})} \times {\hat{\bf n}({\bf r})} \times{\bf E}^i({\bf r})+(1-\alpha)	\hat{{\bf n}}({\bf r}) \times {\bf H}^i({\bf r})=\\-\alpha{\hat{\bf n}({\bf r})} \times\mathcal{T}\circ\mathbf{j}(\rv)+(1-\alpha)(I-\mathcal{K})\circ\mathbf{j}(\rv)
	\end{split}
	\label{eq:cfie}
\end{equation}

where $I$ is the identity operator, $0\le \alpha\le 1$ is a constant weighting factor, and the operators $\mathcal{T}$ and $\mathcal{K}$ are given by:

\begin{equation} 
\begin{split}
\mathcal{T}\circ&\mathbf{j}(\rv) =- \hat{\bf n}({\bf r}) \times \Biggl [ \frac{jk\eta_0}{4 \pi} \int_\Omega d{\bf r}' g(\rv,\rv') {\bf j} ({\bf r}') \\
&+ \frac{j\eta_0}{4 \pi k} \int_\Omega d {\bf r}' \nabla\nabla g(\rv,\rv') \cdot {\bf j} ({\bf r}') \Biggr ] \\
\mathcal{K}\circ\mathbf{j}(\rv) = &\bf \frac{1}{4 \pi} \hat{{\bf n}}({\bf r}) \times  \int_\Omega d{\bf r}' \nabla g(\rv,\rv') \times {\bf j} ({\bf r}')
\end{split}
	\label{eq:ie1}
\end{equation}

Here, $\mb{j}(\rv)=\uvec{n}(\rv)\times\brcs{\mb{H}^i(\rv)+\mb{H}^{s}(\rv)}$ is the induced surface current, $g(\rv,\rv')$ is the free-space Helmholtz Green's function, $k$ is the propagation constant, and $\eta_0$ is the intrinsic impedance of free space.  An $e^{j\omega t}$ dependence is assumed and suppressed.  To construct a moment method system, the surface current is discretized in terms of a set of basis functions as $\mb{j}(\rv)=\sum_{n=1}^{N_s} a_n \mb{f}_n(\rv)$ and the discretized operators in \refp{eq:ie1} are tested with a set of functions $\lbrace \mb{f}_m(\rv)\rbrace,m\in[1,N_s]$ in the usual fashion.
\section{Formulation}
\label{s:Form}
Although many details of the construction of GMM for piecewise flat tessellation and scalar acoustic equations are given in \cite{Nair2011c} and \cite{Nair2012b}, we include a detailed discussion here because these techniques have not been developed for electromagnetic integral equations on smooth surfaces, and because the present method differs in several respects from previous work.  

The GMM is constructed via a decomposition of the manifold $\Omega$ into overlapping subdomains, termed ``patches'', and denoted by the set $\brcs{\Omega_i}$.  Each patch $\Omega_i$ comprises four elements: 1) a set of nodes $\mathcal{N}_i$, 2) a local geometrical parameterization $G_i$ and associated projection plane $\Gamma_i$, 3) a partition of unity $\psi_i$, and 4) a local approximation function space, $\brcs{\mb{f}_k(\rv)}$.  In the following section, we first develop patch definitions and define the partitions of unity, which decouple geometry and approximation functions spaces on neighboring patches.  We then describe in some detail the design of the local geometry descriptions and basis sets.  The set of patches $\brcs{\Omega_i}$ may be constructed in several ways depending on the level of a priori information that is known about the surface.  We employ the following approach, which only requires an oriented point cloud as a starting point. 

\subsection{Neighborhoods from Point Clouds}
\label{ss:Nbhds}
To begin, we require only that $\Omega$ be sampled with a discrete set of $N_m$ nodes $\mathcal{N}\doteq \brcs{n_i}, i=1,2,3,\dots,N_m$ residing in $\mathbb{R}^3$ and equipped with a connectivity map in the sense of nearest neighbors (note that standard simplicial tessellations implicitly contain this information).  
Denote by  $\mathcal{N}_i$ the neighborhood of the $i$th node, which is initially defined as the set containing $n_i$ and the set of its nearest neighbors $\overline{\mathcal{N}}_i \doteq \mathcal{N}_i/n_i$ as specified by the connectivity map.   
Additionally, we require an orientation, specified by a normal defined at each $n_i$.  These normals may be a priori provided, or may be obtained using a classical normal estimation routine, e.g. \cite{Hoppe1992}.  In the case where a node occurs at a geometrical singularity such as a tip or corner, the normal is not uniquely defined; in GMM such nodes are flagged as ``singular'' and are handled using a modified algorithm, which is discussed in a later section.  For the remainder of this section, we assume that the geometry is free from singularities.

Given the node list $\mathcal{N}$ and corresponding normals, for each $n_i$ we define a subdomain $\hat{\Omega}_i$ centered on $n_i$ and bounded by $\overline{\mathcal{N}}_i$.  
Each $\hat{\Omega}_i$ is termed a ``patch primitive''.  We label the set of patch primitives that neighbor $\hat{\Omega}_i$ as $\hat{\Omega}_{\overline{\mathcal{N}}_i}\doteq \brcs{\hat{\Omega}_j | n_j \in \overline{\mathcal{N}}_i}$, i.e. $\hat{\Omega}_{\overline{\mathcal{N}}_i}$ is the set of all patch primitives associated with nodes that neighbor $n_i$, but not including $\hat{\Omega}_i$.  By definition each $\hat{\Omega}_i$ overlaps with all patches in $\hat{\Omega}_{\overline{\mathcal{N}}_i}$.

Although the initial set of patches $\brcs{\hat{\Omega_i}}$ is defined about every node in $\mathcal{N}$, it is often advantageous to merge neighboring primitives into larger patches, e.g., if the patch primitives are small with respect to the characteristic scale of the problem or form part of the same surface of revolution.  We denote the set of merged patches as $\brcs{\Omega_i}$, with each final patch defined as a union of some subset of patch primitives: $\Omega_i\doteq \cup_{k=1}^{N_k} \hat{\Omega}_k$, with $\hat{\Omega}_k$ an $N_k$-dimensional subset of $\brcs{\hat{\Omega}_i}$. After merging, the final patches may be associated with much larger collections of nodes, with neighborhoods $\mathcal{N}_i$ and patch neighbor sets $\brcs{\Omega_{\overline{\mathcal{N}}_i}}$ updated accordingly.  As patch merging algorithms depend on local geometric characteristics, we defer discussion of a sample merging algorithm until section \ref{ss:alg}.   Because there is no formal distinction between merged patches and patch primitives (they are two instances of the same type of object), we drop the hat notation for the remainder of the paper and use the set $\brcs{\Omega_i}$ to denote the set of all patches for a given scatterer, primitive or otherwise.

\subsection{Partition of Unity}
\begin{figure}
	\begin{center}
		\subcaptionbox{ }
		[.5\linewidth]{\includegraphics[width=.48\linewidth]{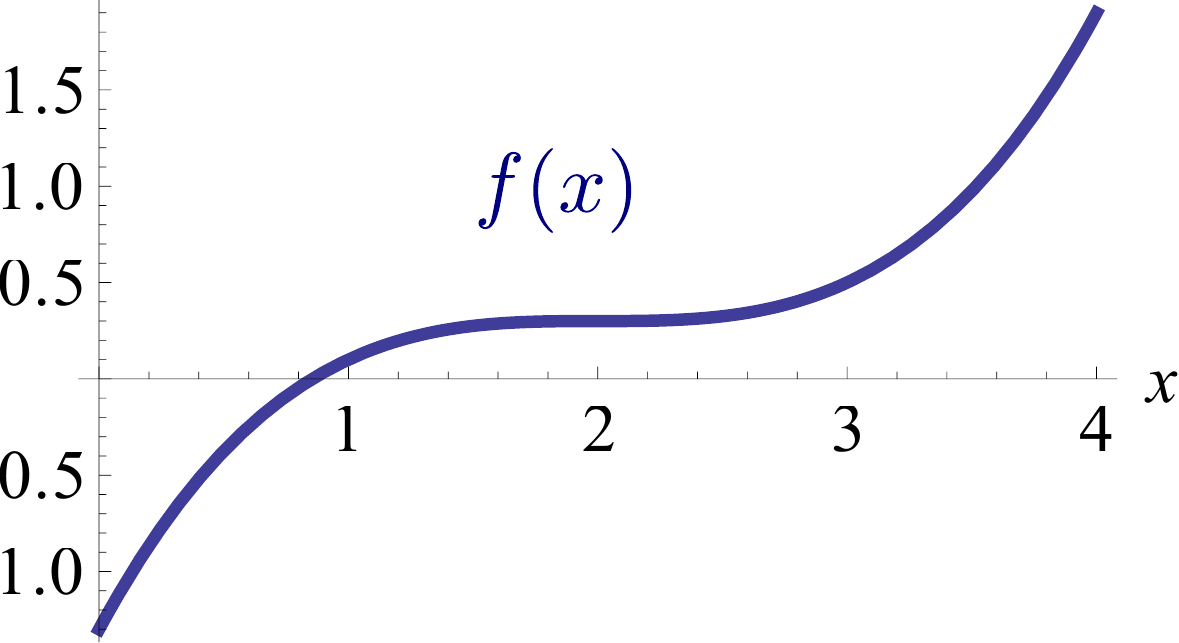}}%
		\subcaptionbox{ }
		[.5\linewidth]{\includegraphics[width=.48\linewidth]{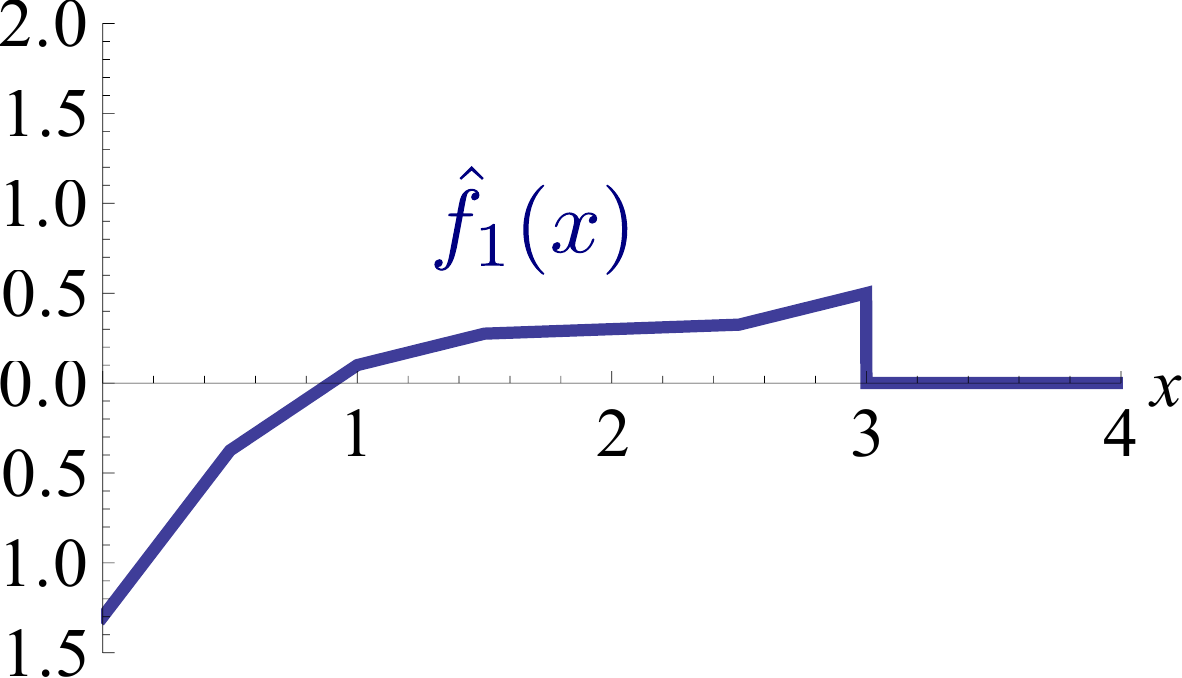}}
		\subcaptionbox{ }
		[.5\linewidth]{\includegraphics[width=.48\linewidth]{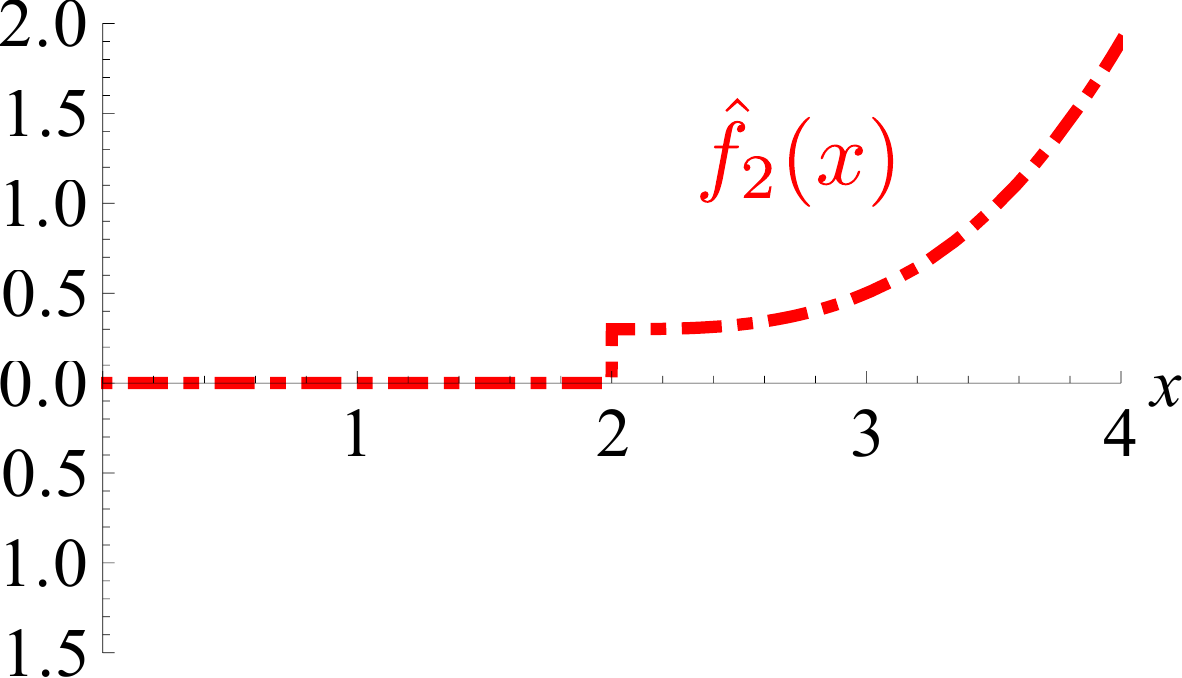}}%
		\subcaptionbox{ }
		[.5\linewidth]{\includegraphics[width=.48\linewidth]{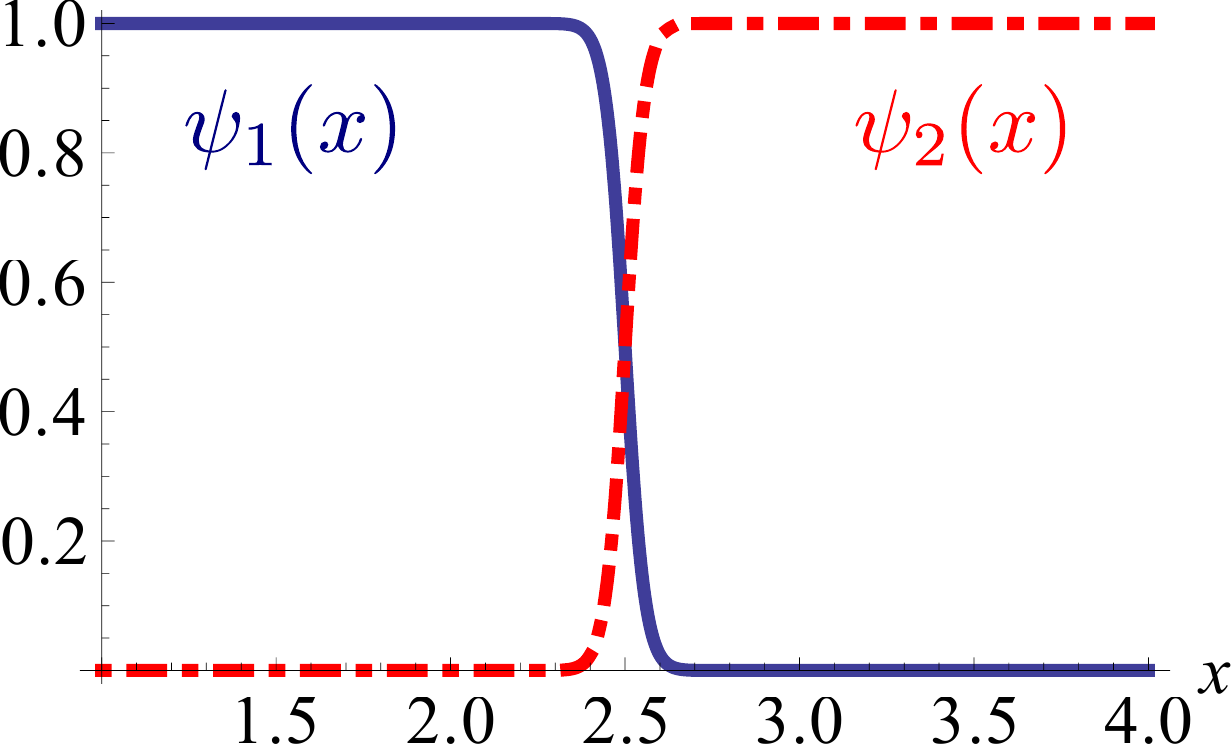}}
		\subcaptionbox{ }
		[.48\linewidth]{\includegraphics[width=.48\linewidth]{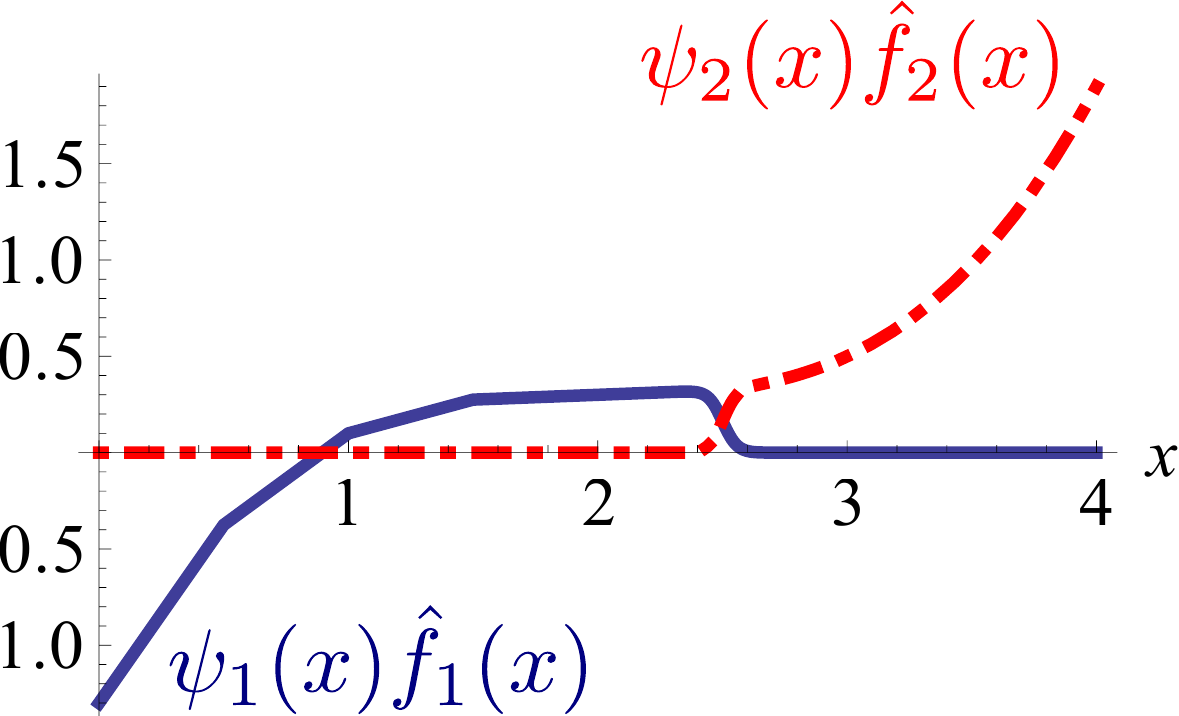}}%
		\subcaptionbox{ }
		[.52\linewidth]{\includegraphics[width=.51\linewidth]{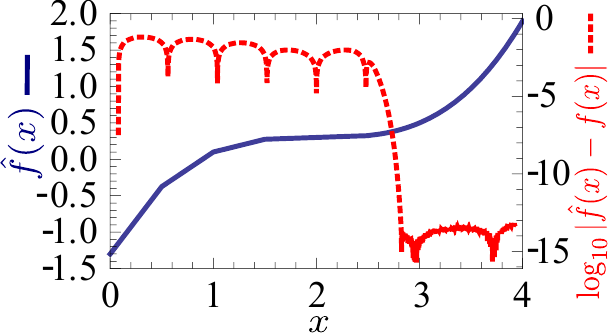}}

		\caption{Blending of approximation spaces on overlapping domains via the partition of unity for a 1D signal.  (a)  A trial function $f(x)=.2(x-2)^3+.3$ on the interval $\Omega:0\le x\le4$.  (b) Interpolation of $f(x)$ using linear interpolatory hat functions $T_i(x)=T(x-.5i)$ on subinterval $\Omega_1: 0\le x\le 3$ such that $\hat{f}_1(x)=\chi_1(x)\sum_i a_i T_i(x)$, $\chi_1(x) = 1$, $x\in\Omega_1$; $0$,  $x \not\in \Omega_1$ .  (c)  Interpolation of $f(x)$ using a third order Legendre polynomial set $P_n(x)$, $n=0,1,2,3$ on subinterval $\Omega_2: 2\le x\le 4$ such that $\hat{f}_2(x)=\chi_2(x)\sum_{n=1}^{3} a_n P_n(x-3)$, $\chi_2(x) = 1$, $x\in\Omega_2$, $0$; $x \not\in \Omega_2$.  (d) Partitions of unity $\psi_1(x)$ and $\psi_2(x)$.  (e) interpolations of $f(x)$ multiplied by partitions of unity as $\hat{f}_1(x)\psi_1(x)$ in $\Omega_1$ and $\hat{f}_2(x)\psi_2(x)$ in $\Omega_2$.  (f) Reconstructed function $\hat{f}(x)=\hat{f}_1(x)\psi_1(x)+\hat{f}_2(x)\psi_2(x)$ (left axis) and absolute reconstruction error (right axis).  Through the action of the partition of unity, the representations $\hat{f}_1(x)$ and $\hat{f}_2(x)$ are smoothly blended in the overlap region, $2\le x\le 3$, and the approximation error transitions smoothly between the linear and Legendre interpolations.}
	\label{fig:PU}
	\end{center}
\end{figure}

Subordinate to each $\Omega_i$, we define a partition of unity function $\psi_i(\rv)$ with $supp\brcs{\psi_i(\rv)}=\Omega_i$ with the property that $\sum_i \psi_i(\rv) = 1 \forall \rv\in \Omega$.  
These partitions of unity serve to decouple neighboring patches, so that any surface function $\phi(\rv),\rv \in \Omega$ may be reconstructed as $\phi(\rv)=\sum_i \psi_i \phi_i(\rv)$, where the subsectional functions $\phi_i(\rv)$ are given by $\phi_i(\rv)=\chi_i \phi(\rv)$, with $\chi_i$ the characteristic function of patch $\Omega_i$. 
In GMM, partition of unity is enforced using an approach based on traditional Shepard Interpolation\cite{Shepard1968}, which takes the form:

\begin{equation}
	\psi_i(\rv)=\frac{\hat{\psi}_i(\rv)}{\hat{\psi}_i(\rv)+\sum_j\hat{\psi}_j(\rv)}
\end{equation}

in which $\hat{\psi}_j(\rv)$ is a local shape function with compact support on patch $\Omega_j$, and $\lbrace \hat{\psi}_j(\rv)\rbrace$ is the set of local shape functions defined on the set $\Omega_{\mathcal{N}_i}$ of patches that neighbor $\Omega_i$.  The choice of local shape function depends on the problem type, and may include simplex type functions, smoothly decaying exponentials, polynomial-based functions, etc.  In the present work, a simplex-based partition of unity is utilized.  Figure \ref{fig:PU} illustrates the blending of basis function spaces via the partition of unity in one dimension.   These concepts extend directly to functions and patches defined on two dimensional manifolds, and also effectively permit blending of different geometry descriptions in the overlap region between patches. 

\subsection{Local Geometry Parameterization}
\label{ss:lGeo}
In this section, we develop methods for constructing local surface geometry descriptions on each patch in $\brcs{\Omega_i}$.  These local surface descriptions are superposed as $\bigcup_i \Omega_i$ to recreate the original geometry.  This scheme admits a wide variety in the choice of local geometry parameterization.   We now describe local geometry construction for both smooth and nonsmooth regions of the scatterer.

From the definition of patches above, each patch $\Omega_i$ contains a set of nodes $\mathcal{N}_i=\brcs{n_k}$ located at positions $\brcs{\rv_k}$ and equipped (in the smooth case) with normals $\brcs{\uvec{n}_k}$.  Local smooth surface parameterization for $\Omega_i$ proceeds as a two-step process.  First, a ``projection plane'' $\Gamma_i$ is defined such that the plane normal $\uvec{n}_i$ is the average of all subordinate node normals: 

\begin{equation}
	\uvec{n}_i=\frac{\sum_k \uvec{n}_k}{|\sum_k \uvec{n}_k|}
	\label{eq:norm}
\end{equation}

The plane $\Gamma_i$ is uniquely defined by its normal and a point $\rv_{c_i}$ (defined with respect to the global origin) through which the plane passes;  generally, this point is taken as the average of the locations of $\brcs{n_k}$, although it may be chosen otherwise depending on the requirements of different geometry parameterizations.  

Once the plane $\Gamma_i$ has been defined, an orthogonal local coordinate system $(u_1,u_2,u_3)$ is constructed as follows.  First, the altitude direction is taken parallel to $\uvec{n}_i$: $\uvec{u}_3=\uvec{n}_i$. The $\uvec{u}_1$ direction is defined in the direction of maximum patch dimension, $\max\lbrace |\rv_{kb}-\rv_{jb}|\rbrace$, where the $b$ subscript indicates a boundary node.  The remaining coordinate direction is then defined as $\uvec{u}_2 = \uvec{u}_3\times\uvec{u}_1$.  The local coordinate system is defined with respect to a local origin located at $\rv_{c_i}$.  

\begin{figure}
	\subcaptionbox{ }[.5\linewidth]{\includegraphics[width=.45\linewidth]{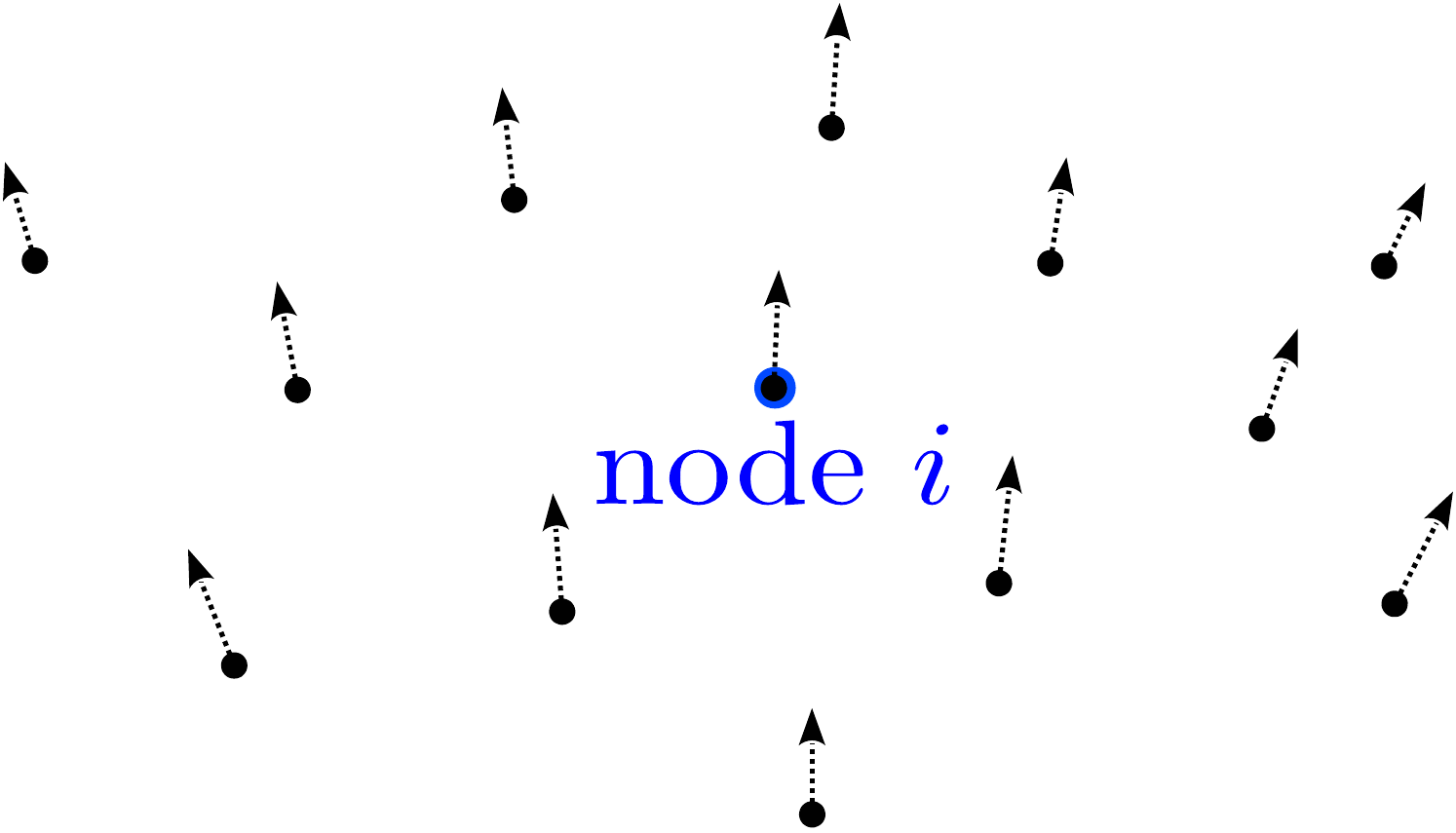}}%
	\subcaptionbox{ }[.5\linewidth]{\includegraphics[width=.45\linewidth]{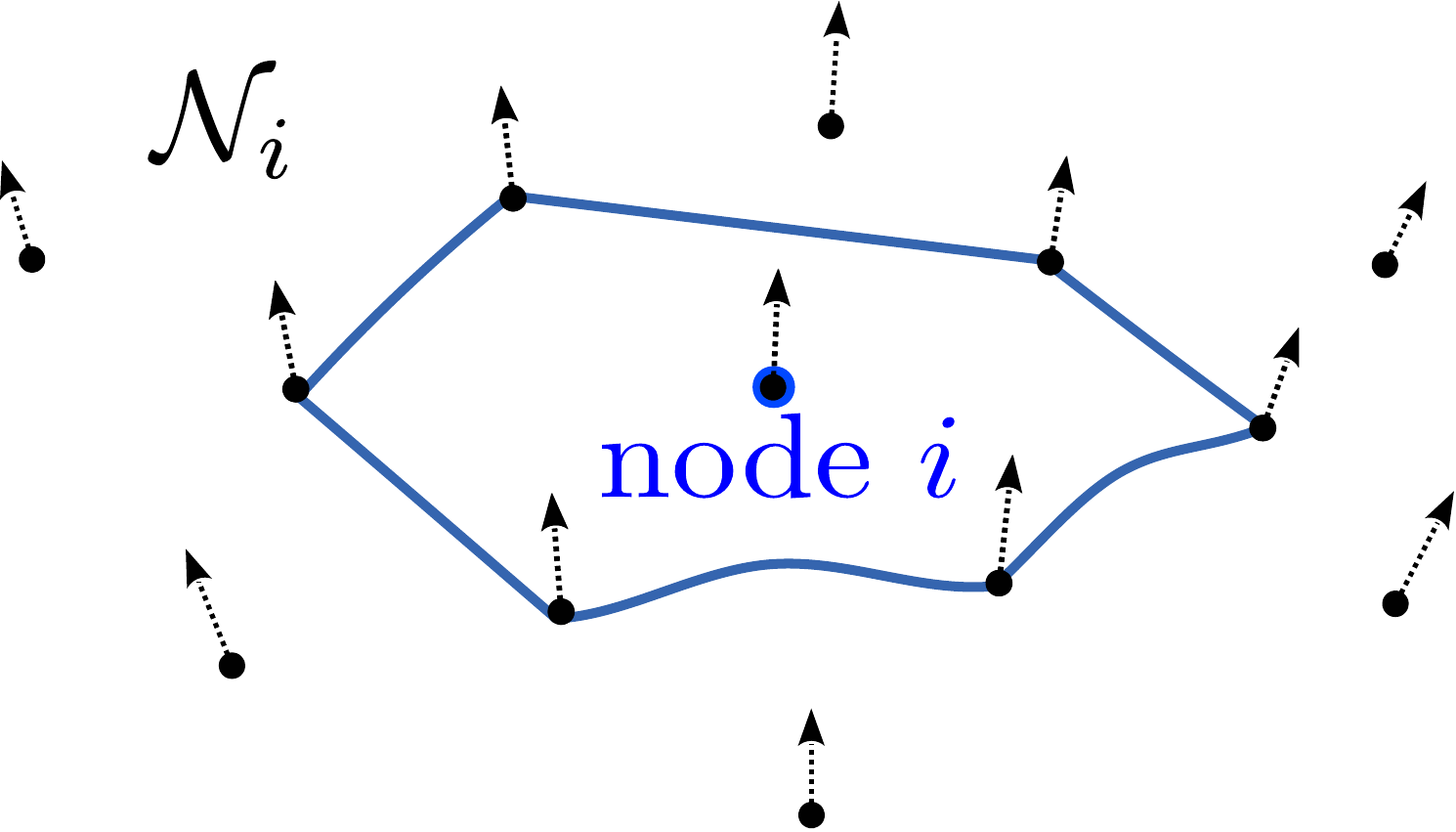}}
	\subcaptionbox{ }[.5\linewidth]{\includegraphics[width=.45\linewidth]{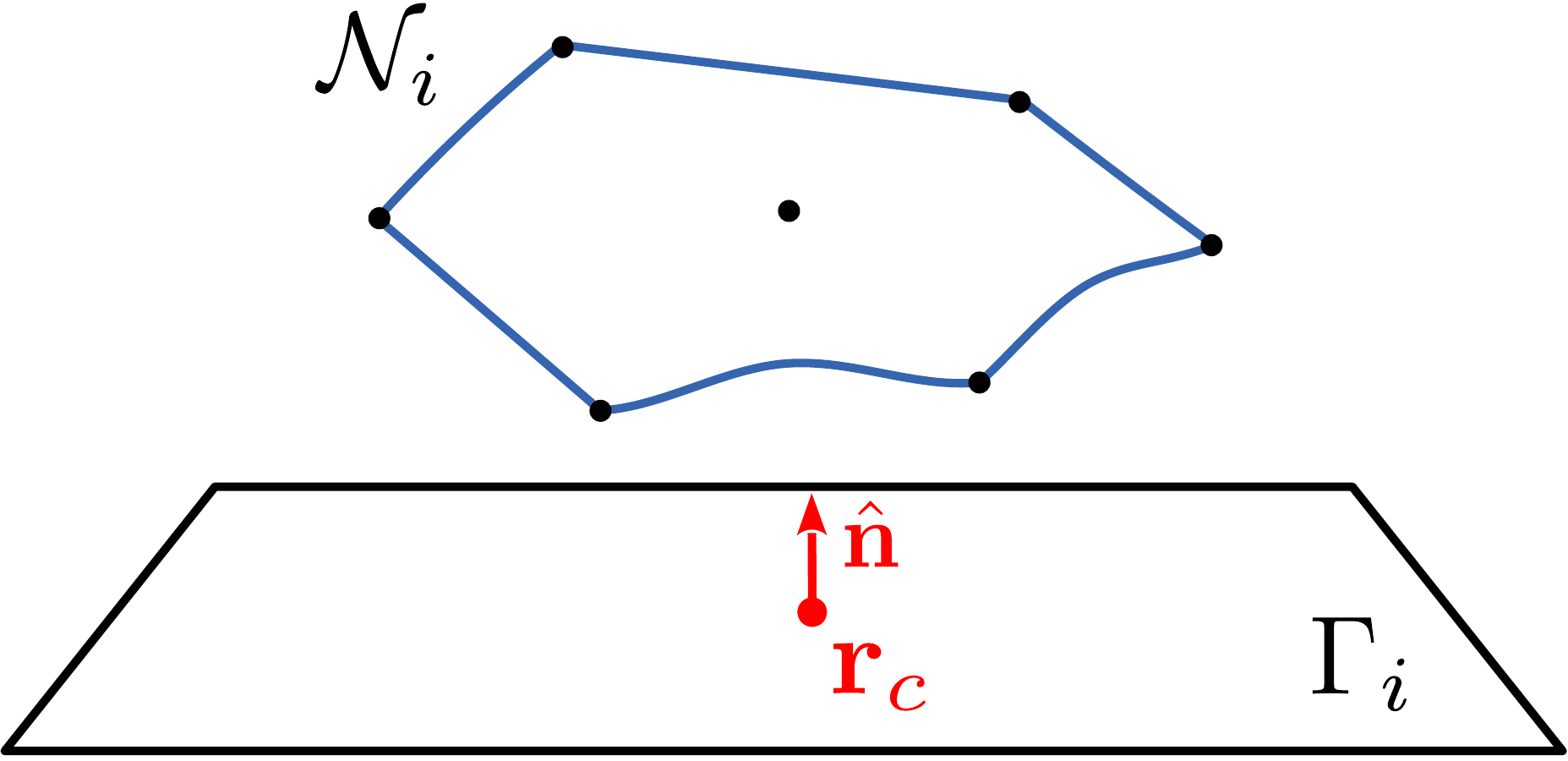}}%
	\subcaptionbox{ }[.5\linewidth]{\includegraphics[width=.45\linewidth]{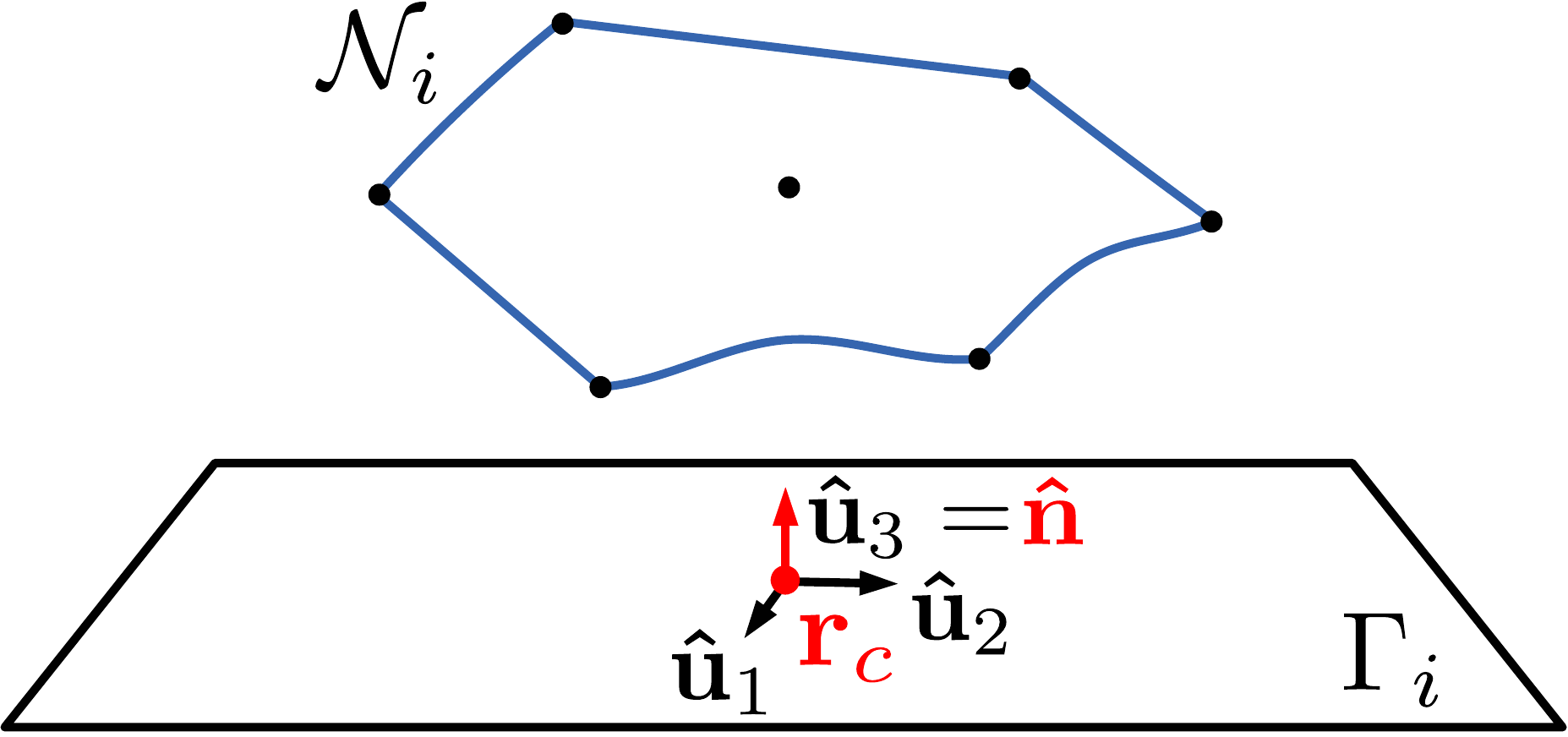}}
	\subcaptionbox{ }[.5\linewidth]{\includegraphics[width=.45\linewidth]{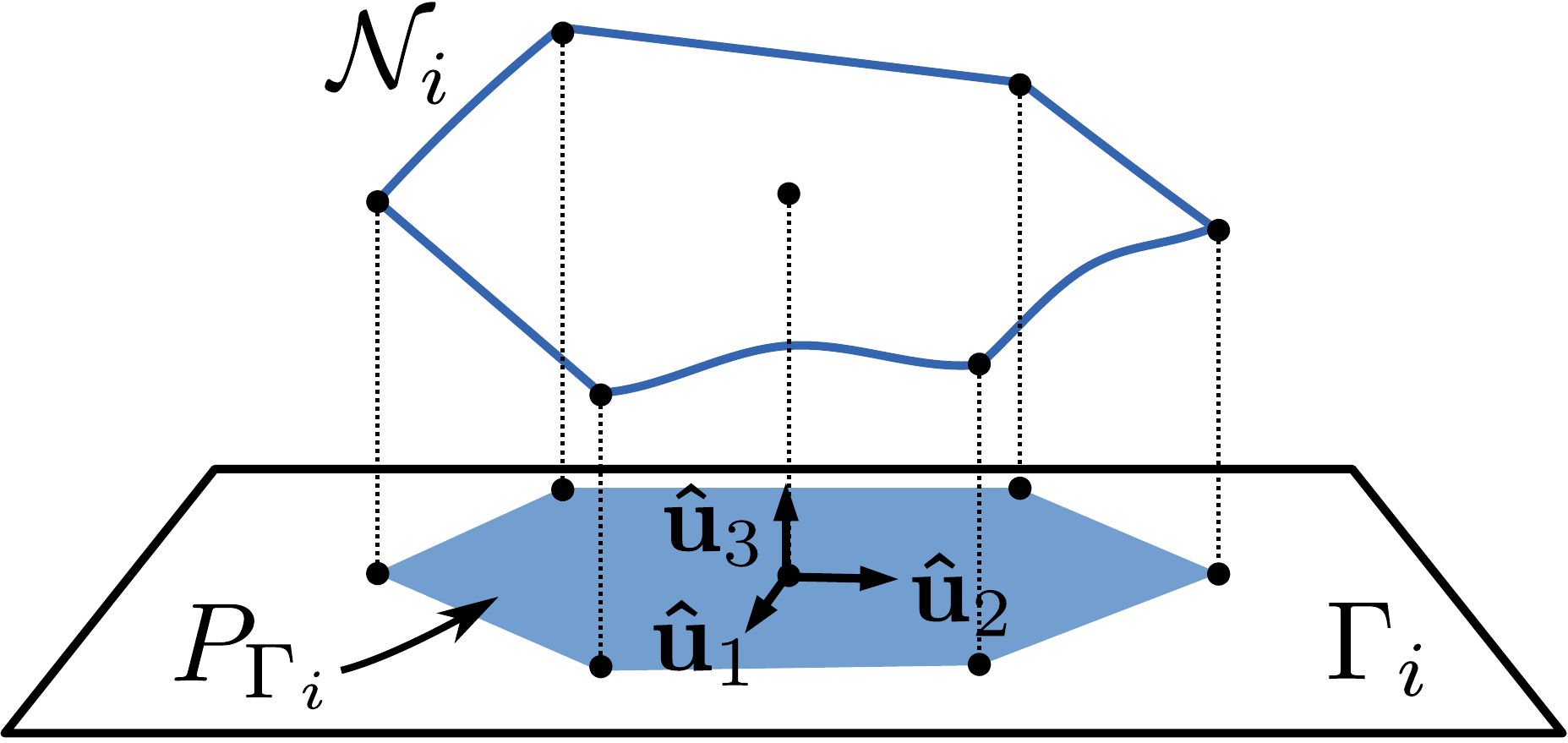}}%
	\subcaptionbox{ }[.5\linewidth]{\includegraphics[width=.45\linewidth]{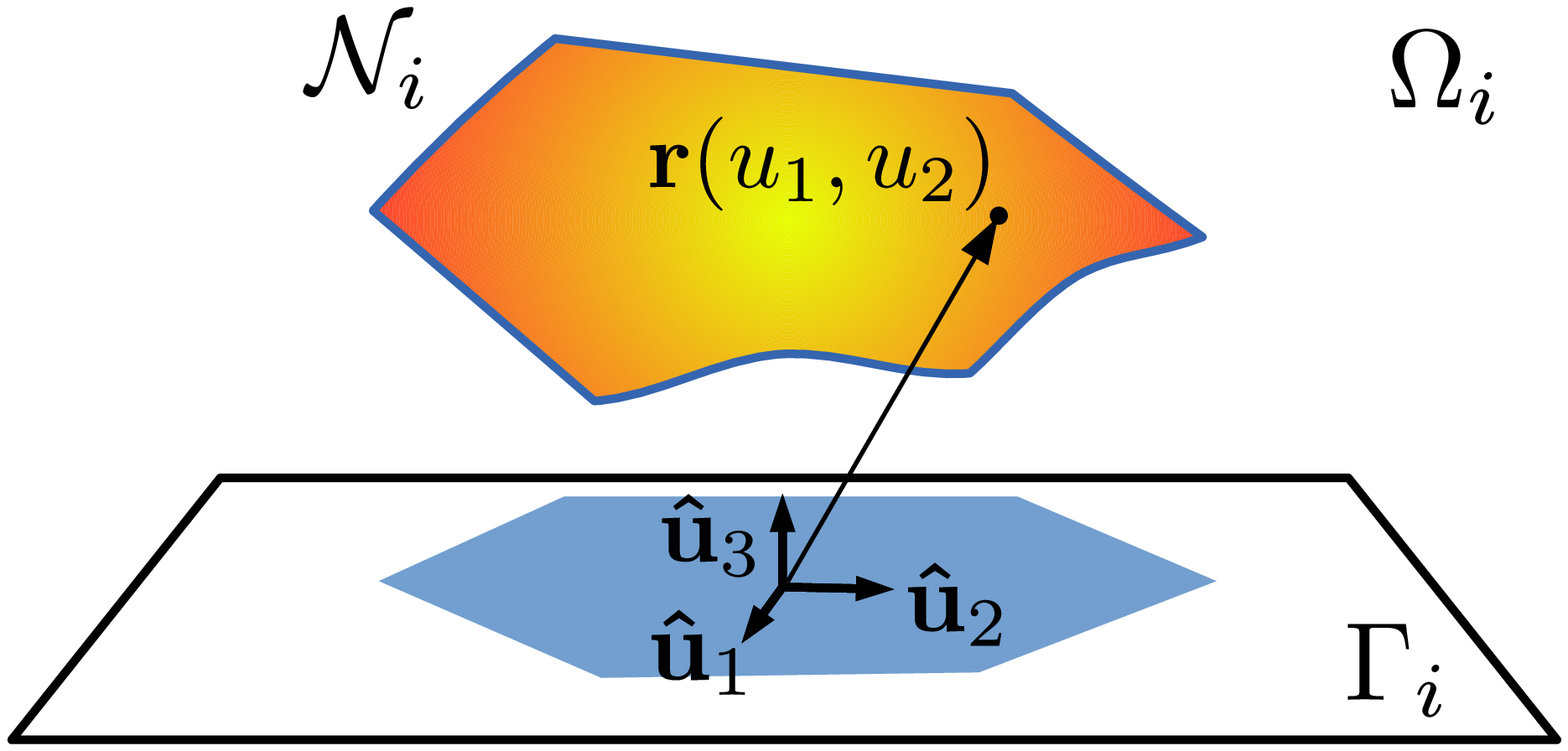}}
	\caption{Smooth Patch Geometry Parameterization Procedure for patch $\Omega_i$:  (note: for clarity, subscripts $i$ are dropped on vector quantities) (a) Start with an oriented point cloud , and (b) define a neighborhood $\mathcal{N}_i$ of nearest neighbors about node $i$.  (c) Define a projection plane $\Gamma_i$ with normal equal to average of all node normals and specified by a point $\rv_c$ that uniquely defines an origin of coordinates.    (d) Construct a local orthogonal $u_1,u_2,u_3$ coordinate system.  (e) Take projection of points in $\mathcal{N}_i$ into $\Gamma_i$ to define the support of the patch in the $u_1,u_2$ plane.  This projection is denoted $P_{\Gamma_i}$. (f) Obtain a smooth parameterization to define the local patch surface.}
	\label{fig:patch}
\end{figure}

Upon establishment of the local coordinate system, a parameterized manifold $\Lambda_i$ passing through the nodes $\brcs{n_k}$ is defined.  This parameterization takes the form:

\begin{equation}
	\begin{split}
		\Lambda_i \doteq & \brcs{\rv\in\mathbb{R}^3 ~ | ~ \rv(u_1,u_2) = \rv_i(u_1,u_2)+\rv_{c_i}},\\
		&\rv_i(u_1,u_2)= u_1\uvec{u}_1+u_2\uvec{u}_2+w_i(u_1,u_2)\uvec{u}_3 
	\end{split}
\end{equation}

where $\rv_i(u_1,u_2)$ indicates a local position vector pointing from the $(u_1,u_2,u_3)$ origin to the parameterized surface.  The domain of the parameterization is restricted to a polygon in the projection plane $\Gamma_i$ that bounds the projection $P_{\Gamma_i}(\mathcal{N}_i)$ of the node set $\mathcal{N}_i$ into $\Gamma_i$:  $u_1,u_2\in P_{\Gamma_i}(\mathcal{N}_i)$.  We denote the type of parameterization on patch $\Omega_i$ by $G_i$.  The altitude function $w_i(u_1,u_2)$ may be taken as any well-behaved function of $(u_1,u_2)$.  The conditions for $w_i(u_1,u_2)$ to be well-behaved are that 1) it is single-valued, 2) its Jacobian is amenable to integration through standard numerical quadrature techniques, and 3) it maintains an appropriate degree of geometrical continuity with neighboring patch descriptions.  As long as these conditions are satisfied, different local geometry parameterizations may be used on different patches, and the choice of parameterization $G_i$ my be matched to local geometric features, e.g. a BoR description for rotationally symmetric surface regions, corner elements for edges and tips, Bezier surfaces, etc.  The mixing of multiple surface descriptions generalizes the work in \cite{Dault2012}, which hybridized polygonal tessellations with smooth polynomial surface descriptions.  Figure \ref{fig:patch} summarizes patch neighborhood and local geometry construction for smooth geometries.  

\subsubsection{Smooth Polynomial Geometry Parameterization}
Although any parameterization satisfying the conditions stated in the previous section may be utilized to define patch surfaces, it is desirable to have a default parameterization scheme for surfaces where no a priori surface description is known apart from the node locations and normals.  To effect this scheme, following \cite{Nair2012} we use a polynomial surface description for the parameterized manifold $\Lambda_i$: 

\begin{equation}
	\rv_i(u_1,u_2)= u_1\uvec{u}_1+u_2\uvec{u}_2+\mathcal{P}_i^{g}(u_1,u_2)\uvec{u}_3
\end{equation}

where $\mathcal{P}_i^{g}(u_1,u_2)$ is as a degree $g$ tensor product of polynomials in $u_1$ and $u_2$:

\begin{equation}
	\mathcal{P}_i^{g}(u_1,u_2)=\sum_{|\alpha|\le g}c_{\alpha} u_1^{\alpha_1} u_2^{\alpha_2}
\end{equation}

Here we have employed standard multi-index notation $\alpha\doteq (\alpha_1,\alpha_2), |\alpha|=\alpha_1+\alpha_2$ for compactness.  The coefficients $c_{\alpha}$ are obtained via least-squares fit.  The geometrical fidelity of $\Lambda_i$ to the original node locations is controllable by the degree $g$ and the tolerance in the least squares algorithm.  By defining $\Lambda_i$ as a polynomial function $\mathcal{P}_i^{g}(u_1,u_2)$, $\Lambda_i$ is ensured to lie the space $C^g$ of $g$-differentiable functions.  This property is important in evaluating field integrals because basis functions in the GMM scheme inherit the differentiability of the geometry parameterization.

\subsubsection{Handling of Geometrical Singularities}
In the case where a patch contains nodes that lie on geometric singularities, a modified geometry parameterization method is employed.  If an analytical description of geometry in the neighborhood of singularity is known, as with a conical tip or straight edge, a custom geometrical description that correctly handles the singularity may be constructed.  Alternatively, a standard triangular tessellation may be defined on the set of nodes adjacent to (and including) the singular nodes.  Appropriate basis sets are then defined on the tessellation that correctly capture the behavior of currents and charges near the singular feature.  Although the design of these functions is outside the scope of this paper, several such basis functions have been developed (the most prevalent being the RWG class of functions) that can be easily incorporated into the GMM framework.

\subsubsection{Vector Calculus on Parametric Surfaces}\label{sss:vc}

Here, we briefly define quantities required to do vector calculus on the parametric surface specified by $\rv(u_1,u_2)$.  Vectors tangential to the surface are expressed as $\mb{t}=a\rv_{u_1}+b\rv_{u_2}$, where $a,b$ are scalar coefficients and the basis vectors $\rv_{u_j}$, $j=1,2$ are defined by $\rv_{u_j}\doteq \partial \rv/\partial u_j$.  Note that the tangent basis vectors are not normalized, and in general do not need to be orthogonal.  For the types of parameterizations used in this paper, the basis vectors are given explicitly by:

\begin{equation}
	\begin{split}
		\rv_{u_1}&=\uvec{u}_1+\dfrac{\partial w(u_1,u_2)}{\partial u_1}\uvec{u}_3\\
		\rv_{u_2}&=\uvec{u}_2+\dfrac{\partial w(u_1,u_2)}{\partial u_2}\uvec{u}_3
	\end{split}
\end{equation}

\begin{figure}
	\centering
	\includegraphics[width=.70\linewidth]{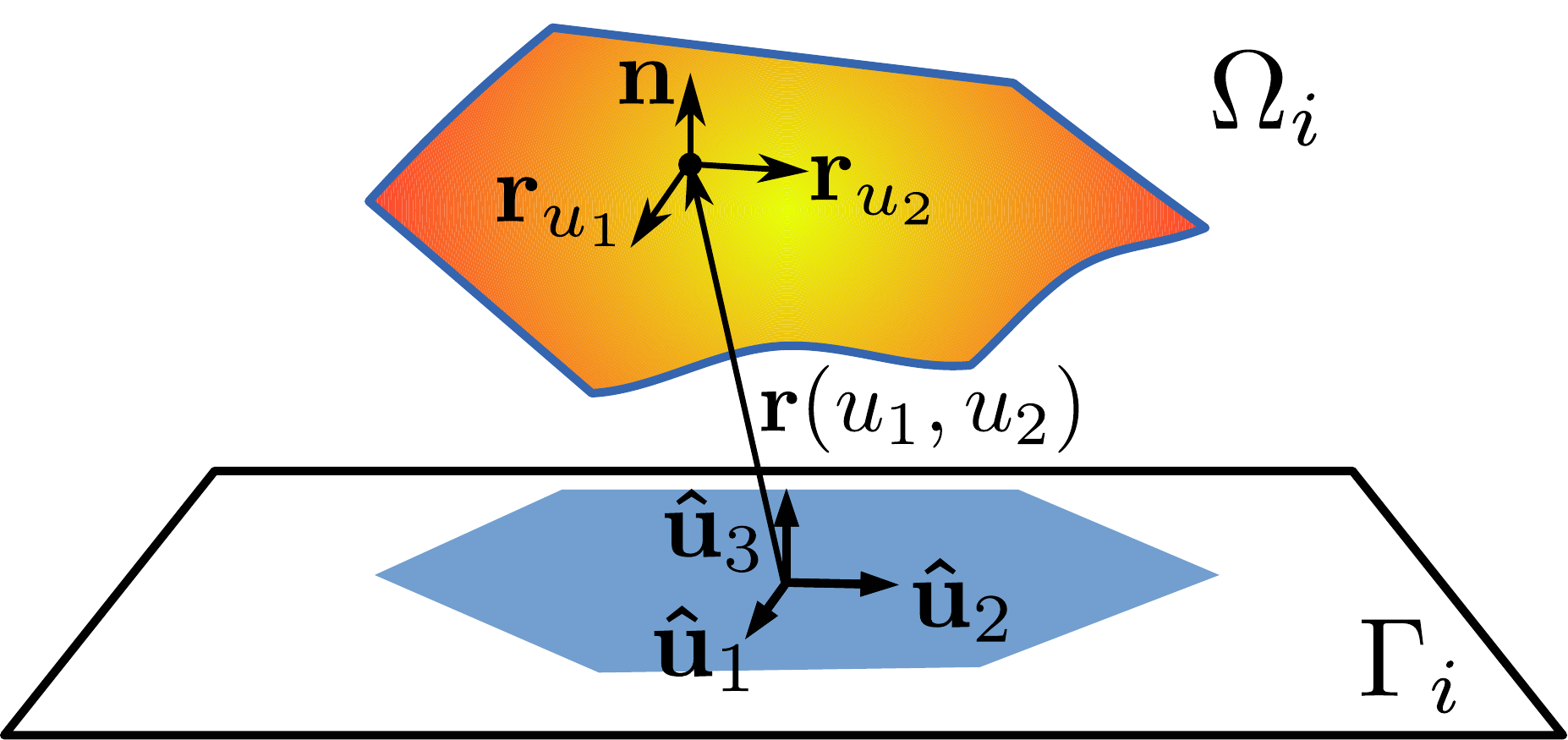}
	\caption{Tangent and normal vectors associated with a parametric patch surface}
	\label{fig:patTan}
\end{figure}

The unit normal is given in terms of these tangent basis vectors as $\uvec{n}=(\rv_{u_1}\times \rv_{u_2})/||\rv_{u_1}\times \rv_{u_2}||$.  Figure {\ref{fig:patTan}} shows $\rv_{u_1},\rv_{u_2}$ and $\uvec{n}$ on the parametric surface.  

For the purposes of integrating and taking derivatives with respect to the parametric coordinates $(u_1,u_2)$, it is necessary to define notions of differential length and area on the parametric surface.  To do so, a transformation matrix called the ``metric tensor'' is defined as:

\begin{equation}
	g=\begin{bmatrix}
		\rv_{u_1}\cdot\rv_{u_1} & \rv_{u_1}\cdot\rv_{u_2}\\
		\rv_{u_2}\cdot\rv_{u_1} & \rv_{u_2}\cdot\rv_{u_2}
	\end{bmatrix}
\end{equation}

Traditionally, components of the metric tensor are indexed using lowered indices $g_{ij}$.  The inverse of the metric tensor $g^{-1}$ is used in expressions involving derivatives, and the components of $g^{-1}$ are referenced using raised indices $g^{ij}$.  The differential surface area element is given by $dS=Jd\mb{u}=Jdu_1du_2$ where the Jacobian is $J=\sqrt{det(g)}$, with $det(g)$ the determinate of the metric tensor.  Finally, the surface divergence of a vector function $\mb{f}(\rv(u_1,u_2))$ on the surface with respect to the parametric coordinates is:

\begin{equation}
	\begin{split}
		\nabla_s \cdot \mb{f}(u_1,u_2) = \sum_{i,j}^2 g^{ij} \dfrac{\partial\mb{f}}{\partial u_i} \cdot \rv_{u_j}
	\end{split}
\end{equation}

This expression is used to compute the divergences of surface currents required in the $\mathcal{T}$ operator of equation {\eqref{eq:ie1}}.  For further development of these concepts in the context of electromagnetics, we refer the reader to {\cite{Peterson2006,Song1995}}.

\subsection{Local Basis Functions} 
Next, we develop local basis function descriptions.  GMM patches may incorporate either entire patch (EP) bases with support over the entire patch,  or sets of sub patch (SP) bases with support smaller than the entire patch. 

\subsubsection{Entire Patch Basis Functions}
The general entire patch vector basis function is given in terms of local coordinates as:

\begin{equation}
	\begin{split}
	\mb{f}_{i,k}(\rv)=&f_{k,1}(u_1,u_2)\rv_{u_1}(u_1,u_2)\\
			&+f_{k,2}(u_1,u_2)\rv_{u_2}(u_1,u_2)
	\end{split}
	\label{eq:vEB}
\end{equation}

so that the unknown surface current on patch $\Omega_i$ is expressed as:

\begin{equation}
	\mb{j}_i(\rv)=\psi_i(\rv)\sum_{k=1}^{N_i}a_k\mb{f}_{i,k}(\rv)
\end{equation}

Here, $\psi_i(\rv)$ is the partition of unity on patch $\Omega_i$, the vectors $\rv_{u_1}$ and $\rv_{u_2}$ are tangent vectors to the surface as defined in section \ref{sss:vc}, and the weighting functions $f_{1,k}(u_1,u_2)$ and $f_{2,k}(u_1,u_2)$ may be chosen from any set of functions that lead to convergent integrals on $\Lambda_i$.  Examples of such choices are hierarchical polynomials, plane waves, body of revolution functions, etc.

\begin{figure}
	\includegraphics[width=.9\linewidth]{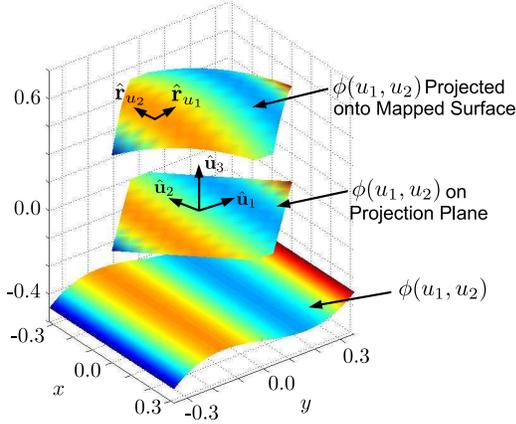}
	\caption{Entire Patch Basis Function}
	\label{fig:EPb}
\end{figure}

To construct EP basis functions, first a canonical minimum bounding shape is defined that encloses the projection $P_{\Gamma_i}$ of the patch node set $\mathcal{N}_i$ onto the local projection plane $\Gamma_i$.  The choice of bounding shape depends on the patch type and shape.  Rectangles are generally employed for mapped polynomial functions, circles for BoR, etc.  Once the appropriate bounding shape is assigned, the component basis functions $f_{k,1}(u_1,u_2)$ and $f_{k,2}(u_1,u_2)$ are defined on its interior and restricted to the domain included in $P_{\Gamma_i}$.  Finally, these basis functions are lifted onto the parametric surface as in \refp{eq:vEB}.  This arrangement is illustrated in figure \ref{fig:EPb}.

\subsubsection{Sub-patch Basis Functions}
\label{ss:SPbf}
GMM can also easily incorporate multiple basis functions on a single patch with sub-patch support, so called sub-patch (SP) bases.  This may be interpreted as defining an EP basis function as $\mathbf{f}(\rv)=\sum_j\mathbf{f}_j(\rv)$, with $\lbrace \mathbf{f}_j(\rv) \rbrace$ a set of approximation functions with support smaller than $supp\brcs{\Omega_i}$. The requirements on SP bases are that they introduce no line charges interior to the patch, and that they possess a normal component at the patch edges.  Due to the partition of unity, the normal component is only required to be finite, and need not satisfy any explicit continuity constraint across patch boundaries.

\begin{figure}
	\includegraphics[width=.9\linewidth]{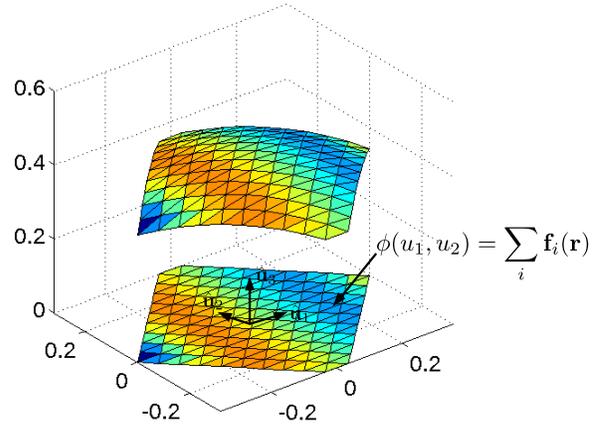}
	\caption{Illustration of sub-patch basis function arrangement with projected PU}
	\label{fig:vSP}
\end{figure}

Sub-patch bases may be defined on the projection plane and projected onto the parameterized surface as with EP bases; alternatively, SP functions may be defined directly on a tessellation of the point cloud as in a traditional Moment Method discretization.  Likewise, the partition of unity is defined either on the projection plane $\Gamma_i$ and then lifted onto the true surface, or is defined directly on the tessellation itself.  Defining sub-patch bases in this way allows most basis function types from existing tessellation-based Moment Method codes to be directly used in the GMM scheme with minimal alteration.  Figure \ref{fig:vSP} shows an example sub-patch tessellation.

	\begin{figure}
		\begin{centering}
		\begin{subfigure}{.5\linewidth}
			{\center \includegraphics[scale=.25]{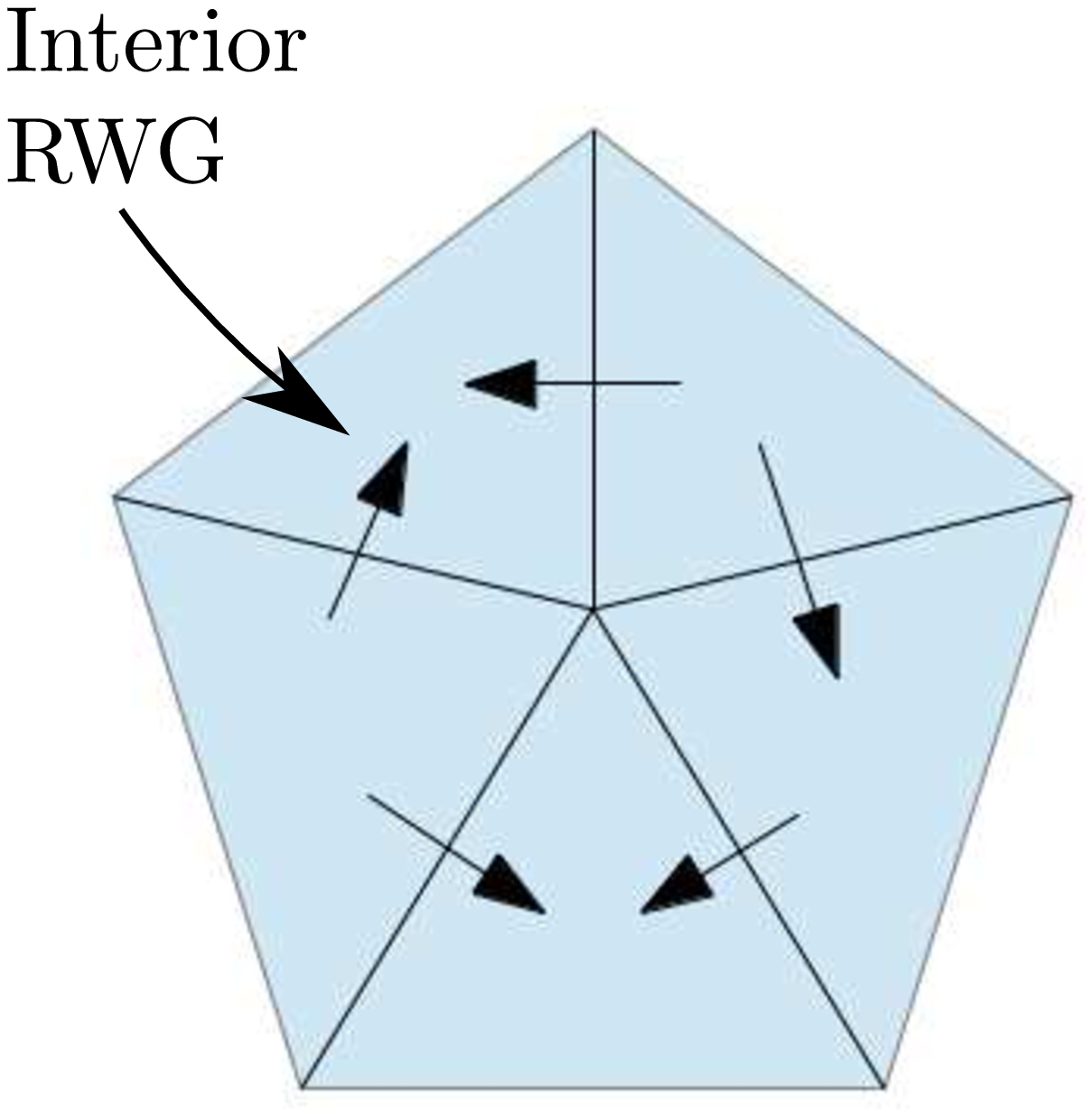}}
			\caption{}
		\end{subfigure}%
		\begin{subfigure}{.5\linewidth}
			{\center \includegraphics[scale=.25]{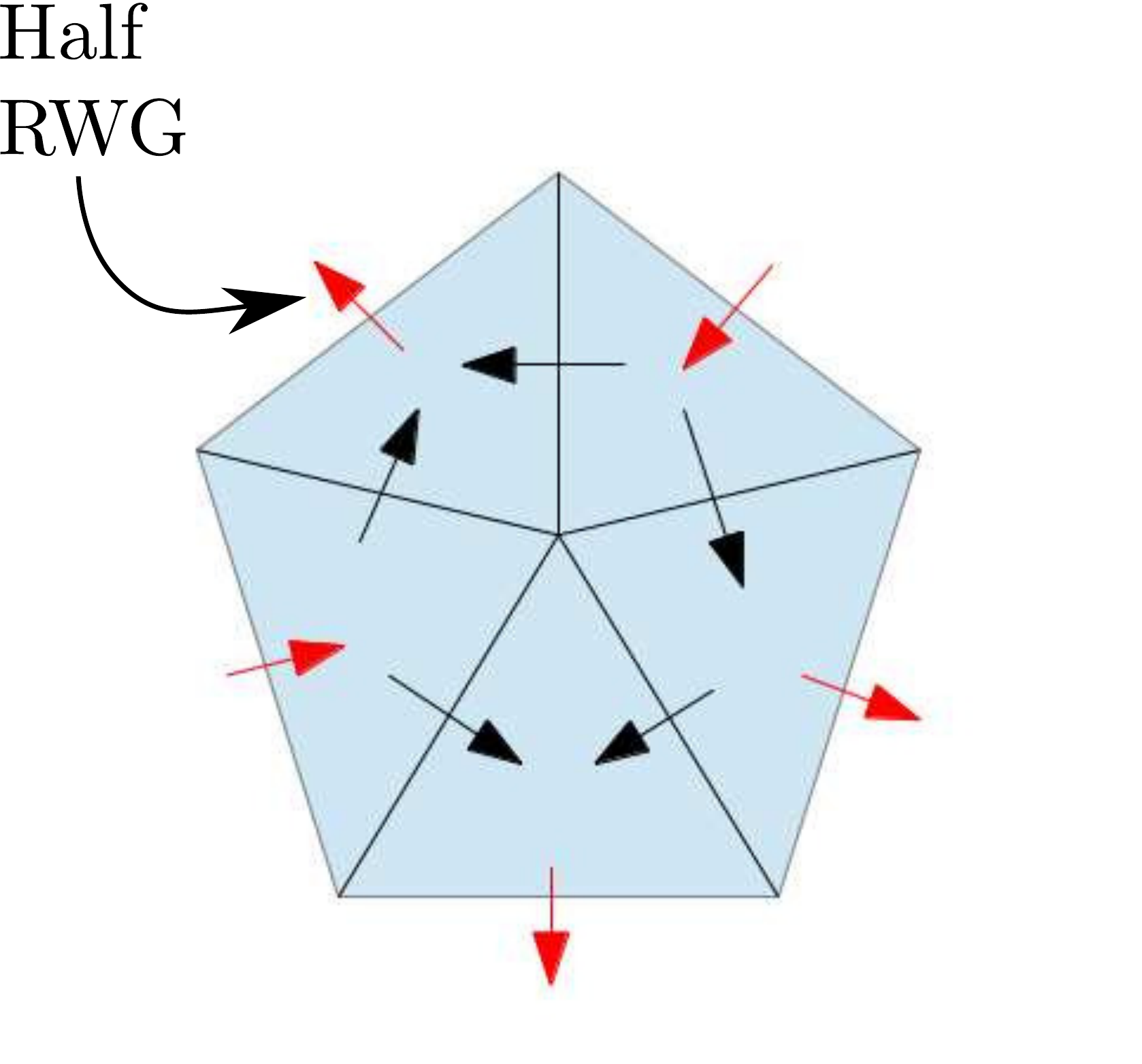}}
			\caption{}
		\end{subfigure}\\
			\begin{center}
		\begin{subfigure}{.5\linewidth}
			\begin{center}
         \begin{tikzpicture}
			\node [inner sep=0pt,above right] 
				{\includegraphics[scale=.2]{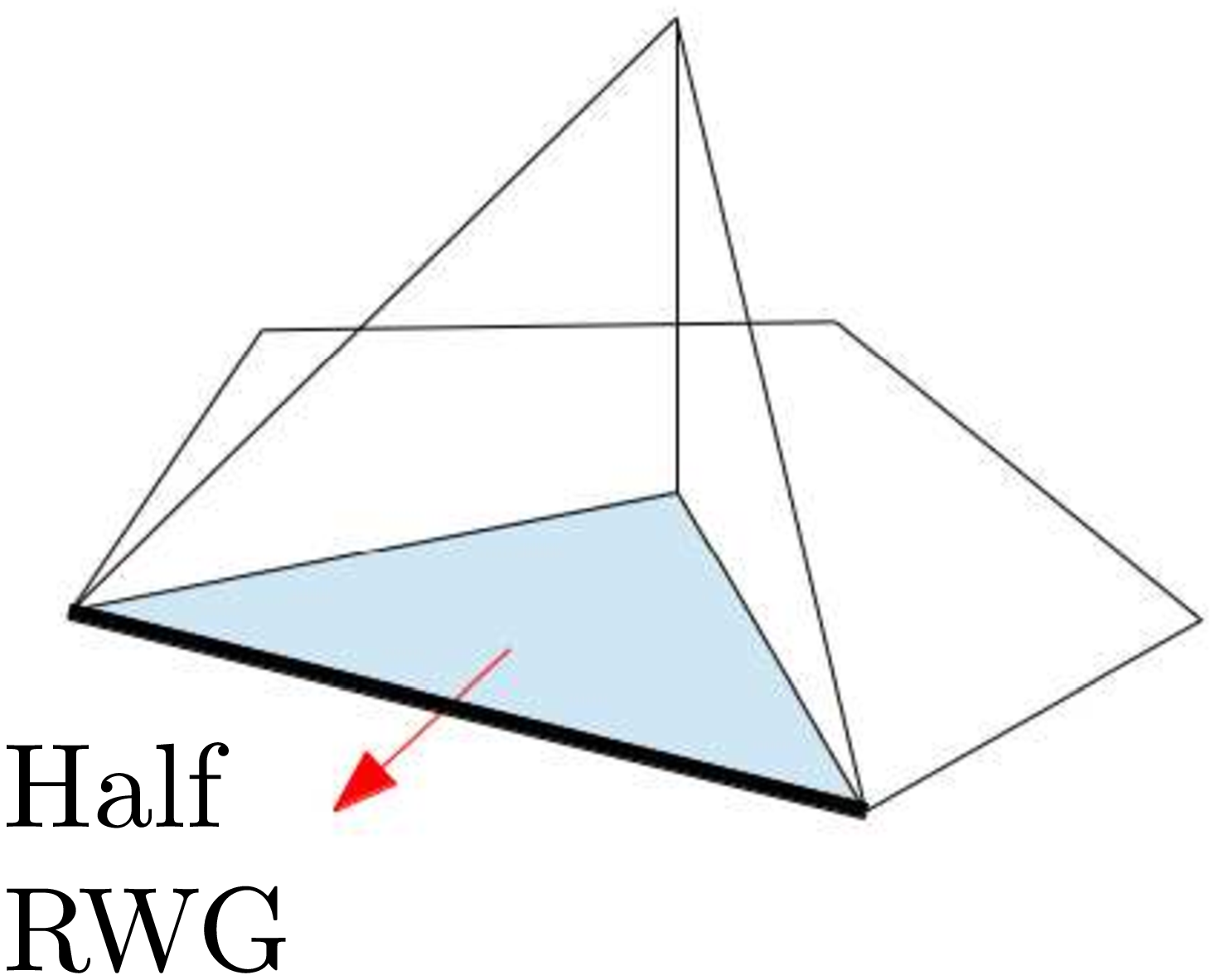}};
			\path (0,1) coordinate (PU);
			\path[->,red] (1,2.3) node[anchor=east]{PU} edge [bend left] (1.4,1.9);
		   \end{tikzpicture}
			\caption{}
			\end{center}
		\end{subfigure}
			\end{center}
			\caption{Design of local RWG basis set with normal continuity across patch boundaries:  (a) Internal RWG basis functions, (b) Added external half-RWG functions to provide a normal component at patch edge, and (c) Simplex-based partition of unity forces the half-RWG function to zero at patch boundary, removing the line charge.}
			\label{fig:RWG_pat}
			\end{centering}
	\end{figure}

	To illustrate the manner in which sub-patch bases are implemented in the GMM scheme, we detail the construction of a tessellated patch with Rao-Wilton-Glisson (RWG) \cite{Rao1982} basis functions.  RWG functions are defined as usual on the tessellated surface, or equivalently, on the projection plane and lifted onto the surface.   A simplicial partition of unity falls naturally into the RWG framework and is defined so that it takes a value of unity in all interior triangles and falls linearly to zero in boundary triangles.  To provide the necessary normal component at the patch edge, a half-RWG function is defined across exterior patch boundaries.  Although half-bases are generally avoided in RWG discretizations because they introduce line charges, this difficulty is avoided in GMM because the PU forces the function to zero at the patch edge.  This property of the PU applies to all GMM basis functions, and renders the basis set on a patch charge neutral, regardless of whether the functions are EP or SP.  Figure \ref{fig:RWG_pat} shows interior and boundary RWG basis functions, and the PU's removal of the boundary line charge.

\subsubsection{Default EP/SP Basis}
As with local geometry descriptions, it is advantageous to have default basis function types that are utilized when a current ansatz is not available or difficult to obtain.  A straightforward choice of an EP basis function is:

\begin{equation}
	\begin{split}
		\mb{f}_1(u_1,u_2)&=\phi_{\bm{\beta}}(u_1,u_2)\rv_{u_1}\\
		\mb{f}_2(u_1,u_2)&=\phi_{\bm{\beta}}(u_1,u_2)\rv_{u_2}
	\end{split}
	\label{eq:lbas}
\end{equation}

where $\phi_{\bm{\beta}}(u_1,u_2)\doteq P_{\beta_1}(u_1)P_{\beta_2}(u_2)$ is a product of two Legendre polynomials of degrees $\beta_1$ and $\beta_2$, and the entire $p$th order representation on patch $\Omega_i$ is constructed as:

\begin{equation}
	\mb{j}_i(\rv)=\psi_i(\rv)\sum_{l=1}^{2}\sum_{|\bm{\beta}|\le p}a_{\bm{\beta},l}\phi_{\bm{\beta}}(u_1,u_2)\rv_{u_l}
	\label{eq:jlbas}
\end{equation}

More complex polynomial bases may be constructed, e.g. one which explicitly enforces a surface Helmholtz decomposition, as in \cite{Nair2011c}.  However, we have found that the minimal tensor product of Legendre polynomials in \refp{eq:jlbas} generally provides better matrix conditioning than other choices of polynomial basis functions for the same order.   Furthermore, since GMM provides current continuity between patches by design, it is not necessary to use the modified Legendre bases or hierarchical bases often employed in higher order mapped methods \cite{Djordjevic2004,Jorgensen2004}.  

With the influence of the partition of unity, the surface current approximation for the default EP basis spans the space:
\begin{equation}
	\begin{split}
		\mb{f}_1(u_1,u_2)\in span\brcs{\psi_i&(\rv)(\phi_{\bm{\beta}}(u_1,u_2)\rv_{u_1})\\ 
		+&\psi_i(\rv)(\phi_{\bm{\beta}}(u_1,u_2)\rv_{u_2})}\\
	|\bm{\beta}|\le p &
	\end{split}
\end{equation}

As described in section \ref{ss:lGeo}, when geometrical singularities such as sharp tips or edges are present, a smooth geometry parameterization and polynomial EP basis cannot be used. Therefore, in these situations, we default to a surface tessellation supporting a SP RWG basis as detailed in \ref{ss:SPbf}.  The properties of these functions have been extensively developed \cite{Rao1982} and will not be repeated here.

\begin{figure}
	\begin{center}
		\subcaptionbox{ }[.6\linewidth]{\includegraphics[width=.6\linewidth]{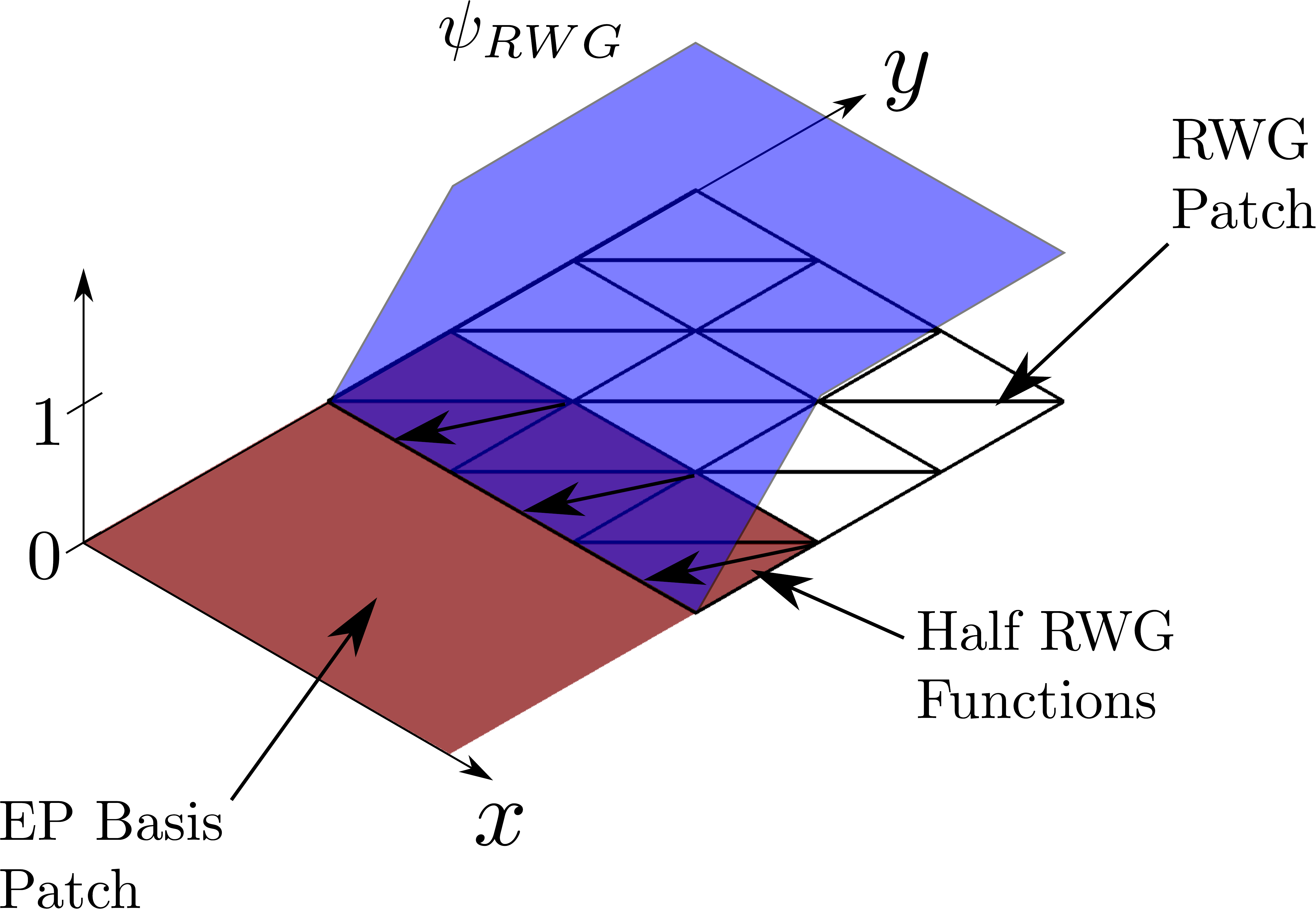}\label{fig:rwgPat}}
		\subcaptionbox{ }[.6\linewidth]{\includegraphics[width=.6\linewidth]{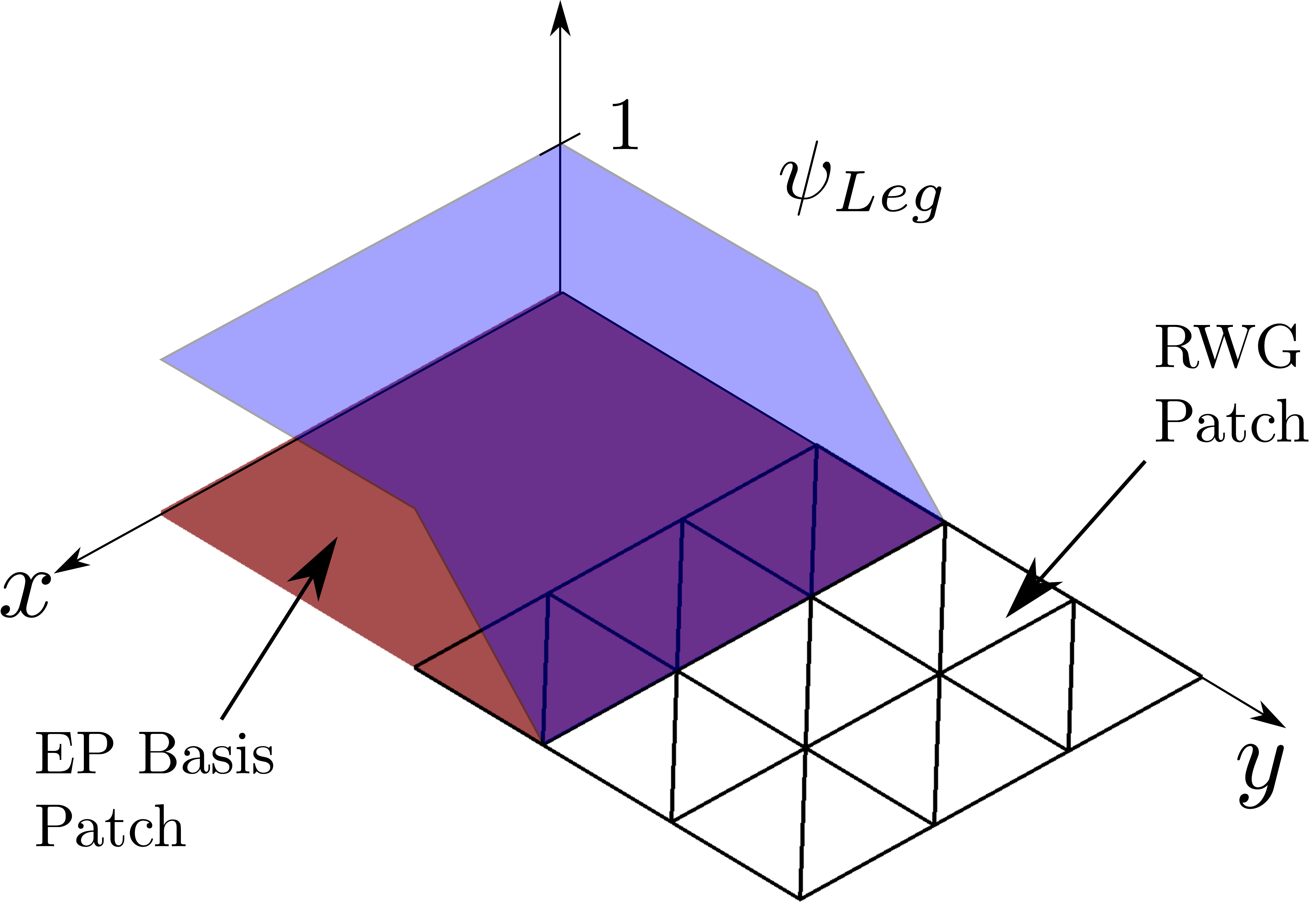}\label{fig:legPat}}
	\end{center}
	\caption{Graphical depiction of partitions of unity for overlapped RWG and Legendre patches. (a) Partition of unity $\psi_{RWG}$ and half-RWG basis functions for an RWG patch.  (b) Partition of unity $\psi_{Leg}$ for a Legendre patch.}
	\label{fig:mixed_bf}
\end{figure}

Finally, we explicitly show how the partitions of unity and half-RWG basis functions are constructed in the case of Legendre (EP-type) and RWG (SP-type) patches that overlap.  Figure {\ref{fig:mixed_bf}}.a shows the partition of unity $\psi_{RWG}=\hat{\psi}_{RWG}/(\hat{\psi}_{RWG}+\hat{\psi}_{Leg})$ and location of half-RWG basis functions for the RWG patch, and Figure {\ref{fig:mixed_bf}}.b shows the partition of unity $\psi_{Leg}=\hat{\psi}_{Leg}/(\hat{\psi}_{RWG}+\hat{\psi}_{Leg})$ on the Legendre patch.  The sub-partition of unity $\hat{\psi}_{RWG}$ on the RWG patch is defined on a triangle-by-triangle basis as:

\begin{equation}
	\hat{\psi}_{RWG}=\begin{cases}
		1,  ~~ n_{bd}=0\\
		\xi, ~~ n_{bd}=1\\
		1-\xi, ~~ n_{bd}=2
	\end{cases}
\end{equation}
where $n_{bd}$ is the number of triangle nodes on the patch boundary (in the overlap region) and $\xi$ is the simplex coordinate associated with the exterior node if $n_{bd}=1$ or the interior node if $n_{bd}=2$.  A similar definition may be used to construct $\hat{\psi}_{Leg}$.

\section{Implementation Details}

In this section, we discuss some practical details of evaluating GMM matrix elements and give sample algorithms for patch merging and basis function assignment.  Each of these tasks is nontrivial, and investigations of the most efficient and effective methods for specific types of problems are subjects of ongoing investigation.  The integration methodology presented is sufficiently general that it may be used for most GMM integrations, although more specialized rules could be developed for specific basis function types.  The integrations and algorithms developed here are used to generate all of the results in section \ref{s:Res}.

\subsection{Matrix Elements}

\begin{figure}
	\begin{subfigure}{.5\linewidth}
			\includegraphics[width=\linewidth]{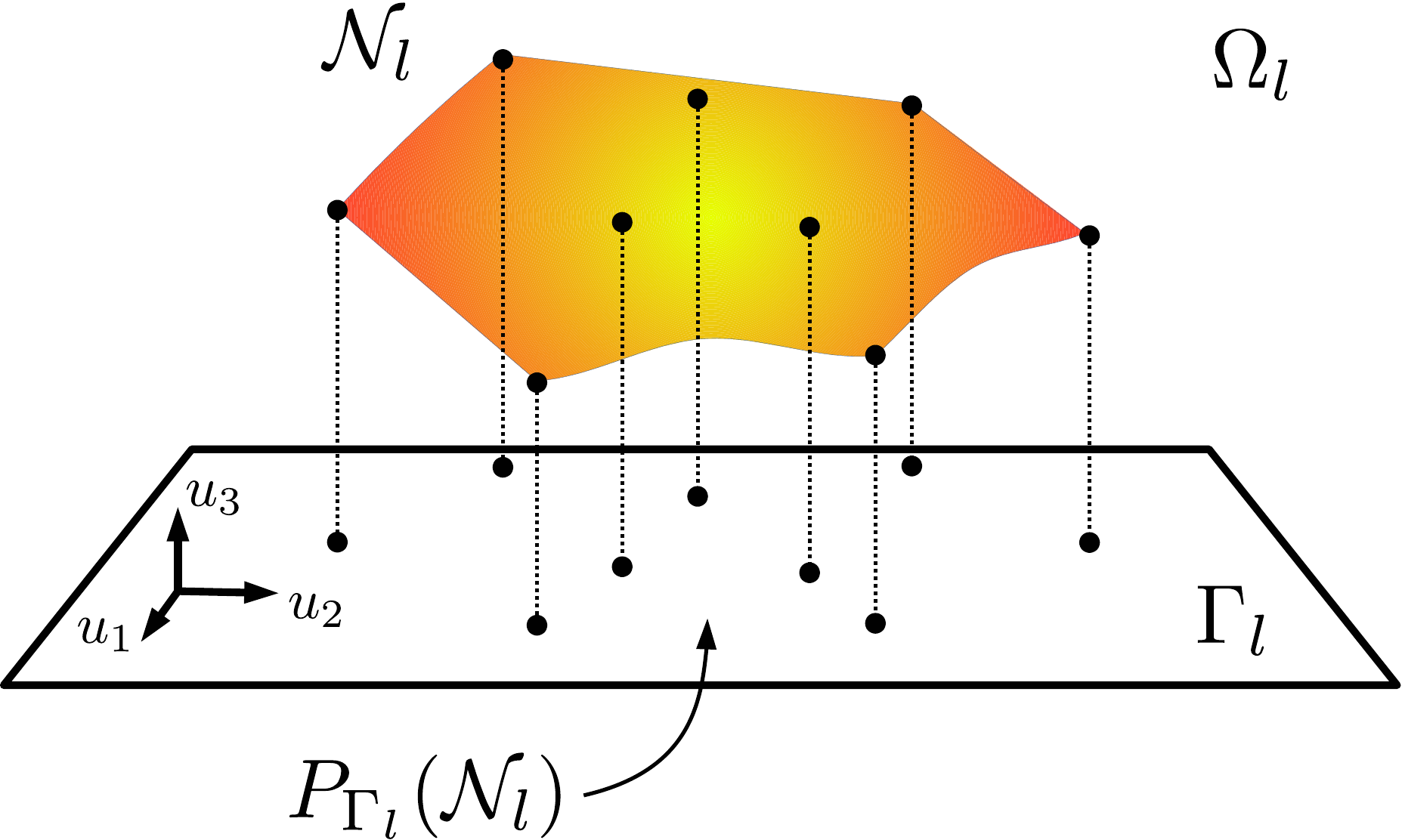}
		\label{sfig:pat_int1}
	\end{subfigure}%
	\hfill
	\begin{subfigure}{.45\linewidth}
			\includegraphics[width=\linewidth]{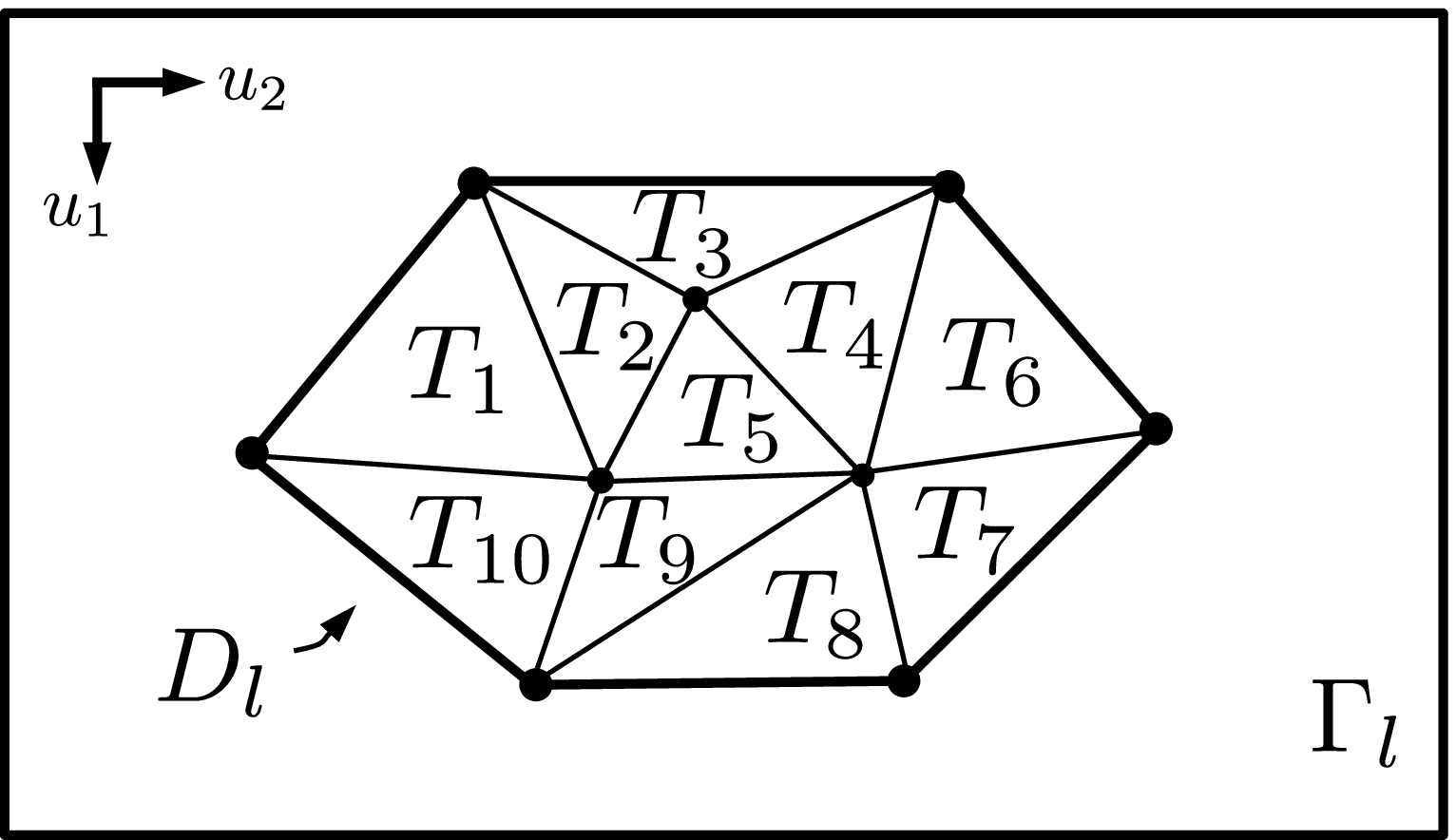}
		\label{sfig:pat_int2}
	\end{subfigure}
	\caption{(a) Graphical representation of $P_{\Gamma_l}(\mathcal{N}_l)$ and (b) construction of triangular integration subdomains. The domain of integration is $D_l=\cup_i T_i$.}
	\label{fig:pat_int}
\end{figure}

We devote this section to a brief discussion of the practical evaluation of GMM matrix elements.  Using the surface parameterizations from Section \ref{ss:lGeo} and the differential geometry quantities from Section \ref{sss:vc}, the tested scattered electric field portion of \refp{eq:ie1} becomes :

\begin{equation} 
	\label{eq:ieuv}
\begin{split}
	-  &\frac{jk\eta_0}{4 \pi} \int_{D_i} d\mb{u}_i J_i(\mb{u}_i) \psi_i(\mb{u}_i)\mb{f}_{i,m}(\mb{u}_i) \\
	\cdot \int_{D_j} & d\mb{u}_j' J_j(\mb{u}_j') g(\rv_{i}(\mb{u}_i),\rv_{j}(\mb{u}_j')) \psi_j(\mb{u}_j)\mb{f}_{j,n}(\mb{u}_j')) \\
	&+ \frac{j\eta_0}{4 \pi k} \int_{D_i} d\mb{u}_i J_i(\mb{u}_i) \nabla_s \cdot (\psi_i(\mb{u}_i)\mb{f}_{i,m}(\mb{u}_i)) \\
	&\int_{D_j} d\mb{u}_j' J_j(\mb{u}_j') g(\rv_{i}(\mb{u}_i),\rv_{j}'(\mb{u}_j')) \nabla_s' \cdot (\psi_j(\mb{u}_j)\mb{f}_{j,n}(\mb{u}_j'))
\end{split}
\end{equation}

where we have employed the standard derivative transfers in the $\Phi$ portion of the $\mathcal{T}$ operator.  The tested magnetic field integral equation portion of \refp{eq:ie1} is:

\begin{equation} 
	\label{eq:mfie_uv}
\begin{split}
	\frac{1}{4 \pi} & \int_{D_i} d\mb{u}_i J_i(\mb{u}_i) \psi_i(\mb{u}_i)\mb{f}_{i,m}(\mb{u}_i)\psi_j(\mb{u}_j)\mb{f}_{j,n}(\mb{u}_j) -\\
	\int_{D_i} & d\mb{u}_i J_i(\mb{u}_i) \mb{f}_{i,m}(\mb{u}_i)\cdot \uvec{n}(\rv(\mb{u}_i))\times\\
	\int_{D_j} & d\mb{u}_j' J_j(\mb{u}_j') \nabla g(\rv_{i}(\mb{u}_i),\rv_{j}(\mb{u}_j')) \times \psi_j(\mb{u}_j)\mb{f}_{j,n}(\mb{u}_j')
\end{split}
\end{equation}

were $\mb{u}_{l}=(u_1,u_2)\in\Gamma_{l}$ denotes the local transverse coordinates for patch $l$, $J_l$ is the Jacobian as defined in section \ref{sss:vc}, and $D_l\in\Gamma_l$ is the domain of integration in the projection plane.  If the projection $P_{\Gamma_l}(\mathcal{N}_l)$ of the nodes in $\mathcal{N}_l$ is a canonical integration domain, e.g. a regular polygon, then an appropriate quadrature rule can be used to integrate over the entire projection. For instance, radial/angular rules are used for BoR patches.  In the general case where the projection is not bounded by a canonical shape,  we perform a triangulation of $P_{\Gamma_l}$, e.g. using Delaunay or another method.   Figure \ref{fig:pat_int} illustrates this process.  Subdividing the domain of integration in this way allows the use of well-developed integration rules for triangles (including singularity rules), and the total integral is the sum of integrations over subtriangles.  It is important to note that not every node in $P_{\Gamma_l}$ needs to be used in defining the triangulation: if the patch is large, some subset of the nodes may be used to lessen the integration cost, as long as the patch boundaries are correctly handled.

\subsection{Automated Patch Construction}
\label{ss:alg}
For large, complex geometries, it is necessary to automate the processes of patch construction including merging, geometry parameterization, and basis function assignment.  The following are two sample algorithms for merging and geometry/basis selection. 

\subsubsection{Algorithm for Patch Merging}
\label{sss:mrgalg}
Using the procedure from section \ref{ss:Nbhds}, patch primitives are assigned in the initial pass. To reduce the number of degrees of freedom in the MoM system, it is often advantageous to merge patches if neighboring primitives share common geometrical features, e.g. are part of the same body of revolution or fall on a sufficiently smooth portion of $\Omega$.  Given the definition of $\Omega_i$ as a collection of nodes, merging and or splitting may be simply done by operating on the sets $\mathcal{N}_i$.  Algorithm \ref{alg:mrg} illustrates a procedure for merging; splitting may be done in a similar fashion.  Criteria for merging are tied both to patch size and patch mean curvature, which is approximated via deviation of normals as shown in Algorithm \ref{alg:mrg}. 

\begin{algorithm}
  \begin{algorithmic}[1]
	 \hrule
	 \vspace{.2cm}
	 \STATE Define maximum normal deviation $\epsilon_n$ tolerance
	 \STATE Define maximum average normal deviation $\epsilon_d$ tolerance
	 \STATE Define maximum patch diameter $D_m$
	 \FOR{each patch primitive $\hat{\Omega}_i\in\lbrace\hat{\Omega}_i\rbrace$}
	 \FOR{each $n_j\in \mathcal{N}_i$ (corresponding to primitive $\hat{\Omega}_j$}
			\IF{$\uvec{n}_j\cdot\uvec{n}_i>(1-\epsilon_n)$}
			\STATE Define candidate merged neighborhood: $\mathcal{N}_l=\mathcal{N}_i\bigcup\mathcal{N}_j$.  Designate the number of nodes in $\mathcal{N}_l$ as $M$.
			\STATE Compute new average patch normal $\uvec{n}_l=\bigl(\sum_{k=1}^{M} \hat{\bf n}_k\bigr )/|\sum_{k=1}^{M} \hat{\bf n}_k|$
				 \STATE Compute merged patch maximum diameter $D=max\lbrace |\rv_{n_a}-\rv_{n_b}|\rbrace,~n_a,n_b \in \mathcal{N}_l$
				 \IF{$D\le D_m$}
				 \STATE Compute average deviation $\overline{\bf \Delta \uvec{n}}_l$ from patch normal $\uvec{n}_l$ as: $\overline{\bf \Delta \uvec{n}}_l \doteq \frac{1}{M}\sum_{k=1}^{M} \hat{\bf n}_l\cdot\hat{\bf n}_k$ 
					 \IF{$\overline{\bf \Delta \uvec{n}}_l\le \epsilon_d$}
						 \STATE \emph{Merge $\hat{\Omega}_i$, $\hat{\Omega}_j$:}
						 \STATE $\mathcal{N}_i\leftarrow\mathcal{N}_l$
						 \STATE $\lbrace\hat{\Omega}_i\rbrace\leftarrow\lbrace\hat{\Omega}_i\rbrace\backslash\hat{\Omega}_j$
					\ENDIF	
				\ENDIF
			\ENDIF
		\ENDFOR
	\ENDFOR
	 \vspace{.2cm}
	 \hrule
	 \vspace{.2cm}
	\end{algorithmic}
  \caption{Automated patch merging} 
  \label{alg:mrg}
\end{algorithm}

At the end of the merging process, the final set of merged patches is designated $\brcs{\Omega_i}$.  Algorithm $\ref{alg:mrg}$ is a simplified version of the actual merging algorithm used in the present GMM code.  In practice, additional higher level merging conditionals are also used that, for instance, favor convex patch boundaries over concave boundaries, attempt to maintain equal patch sizes over the entire geometry, and group patches with similar characteristics, such as edge singularities or small curvature.  

\subsubsection{Automated Geometry and Basis Assignment}

Algorithm \ref{alg:bf} illustrates automated assignment of local geometry descriptions $G$, basis function types $b$ and (where applicable) orders $g$ using using tessellations, BoR surfaces, and polynomial smooth surfaces for geometry parameterization and RWG, BoR, polynomial, and plane wave bases.  Any geometry or basis function class valid for the GMM framework may be included in the algorithm with the proper conditionals.  The present code implementing GMM is written in a highly modularized fashion that facilitates easy addition of new geometry and basis function types. 

\begin{algorithm}
  \begin{algorithmic}[1]
	 \hrule
	 \vspace{.2cm}
	 \STATE Define maximum patch normal deviation $\epsilon_d$ for smooth surface description
	 \STATE Define maximum patch normal deviation for plane wave basis $\epsilon_p$ 
	 \STATE Define minimum patch diameter for plane wave $D_p$
    \FOR{each final patch $\Omega_i$}
	 \STATE Compute average deviation $\overline{\bf \Delta \uvec{n}}_i$ from average patch normal $\uvec{n}_i$ as: $\overline{\bf \Delta \uvec{n}}_i \doteq \frac{1}{M}\sum_{k=1}^{M} \hat{\bf n}_i\cdot\hat{\bf n}_k$
	 \IF {$\overline{\Delta \uvec{n}}_i>\epsilon_d$}
        \STATE $G$: triangular tessellation  
		  \STATE $b$: RWG Basis
	  \ELSIF {Rotationally Symmetric Patch}
		  \STATE $G$: Axisymmetric BoR representation 
		  \STATE $b$: BoR Basis
	  \ELSE 
    	\STATE $G$: Local smooth approximation to $\Omega_i$. 
		\IF{$\overline{\Delta \uvec{n}_i}<\epsilon_p$ \&\& $Diam\brcs{\Omega_i}> D_p$}
			\STATE $b$: Plane Wave Basis
		\ELSE
			\STATE $b$: Polynomial Basis of order $g$
		\ENDIF	
    \ENDIF
	 \ENDFOR
	 \vspace{.2cm}
	 \hrule
	 \vspace{.2cm}
\end{algorithmic}
  \caption{Automated geometry and basis assignment} 
  \label{alg:bf}
\end{algorithm}
Algorithms \ref{alg:mrg} and \ref{alg:bf} are used to discretize all of the scatterers in the results section.
\section{Results}
\label{s:Res}
In this section, we provide results that illustrate hybridization of geometry and basis types and verify the accuracy and flexibility of the method.  
First, we utilize a representation test to demonstrate the hybridization of multiple basis types in the reconstruction of an analytical function.  We then provide several scattering results, each with a different mixture of basis function types and orders defined on patches with varying local geometry descriptions.  Throughout the results section, the algorithms detailed in Section \ref{ss:alg} are used to automatically assign local geometrical descriptions and surface current approximation spaces that are matched to solution ansatzes for the continuous problem.  Doing so gives accurate solutions with significant reductions in the numbers of degrees of freedom relative to a reference CFIE-RWG code.

				\begin{figure}
						\subcaptionbox{ }[.45\linewidth]{\includegraphics[width=.45\linewidth]{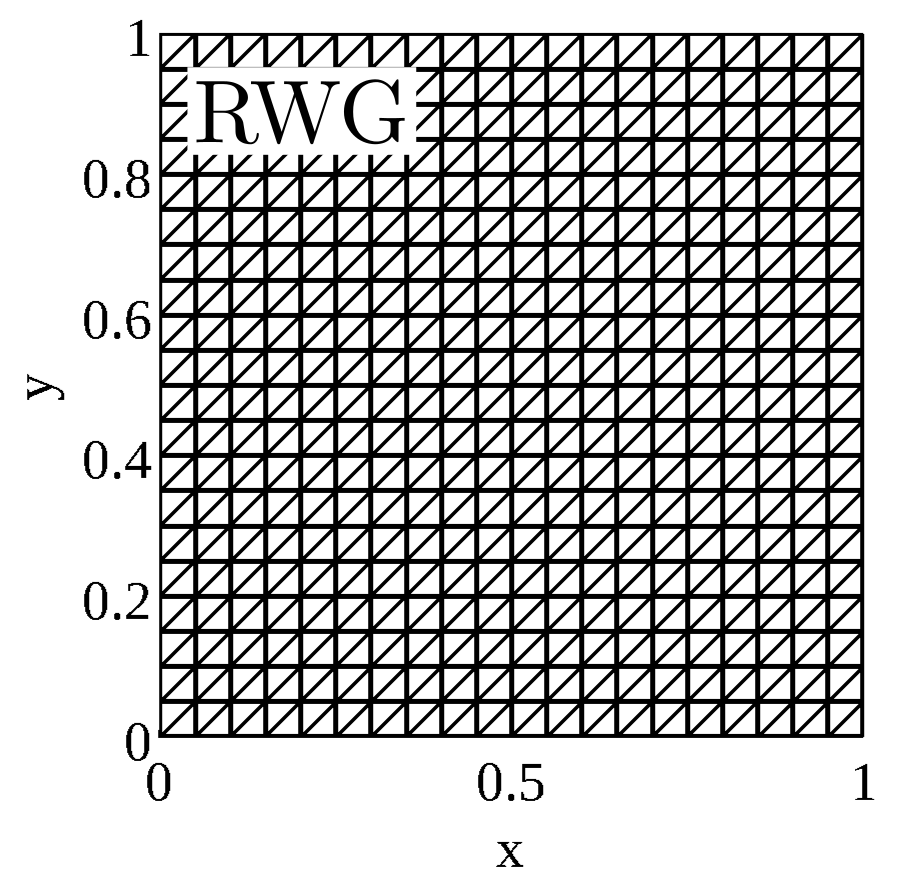}\label{fig:rwgDisc}}%
					\subcaptionbox{ }[.45\linewidth]{\includegraphics[width=.55\linewidth]{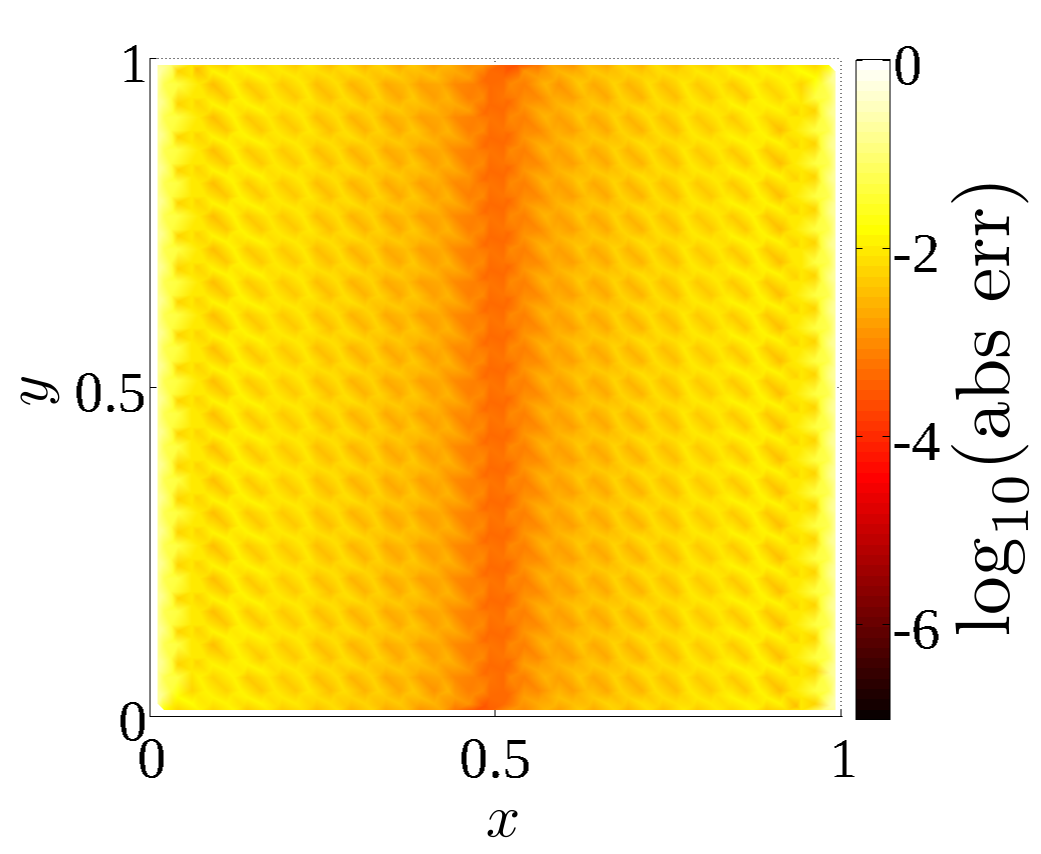}}
					\subcaptionbox{ }[.45\linewidth]{\includegraphics[width=.45\linewidth]{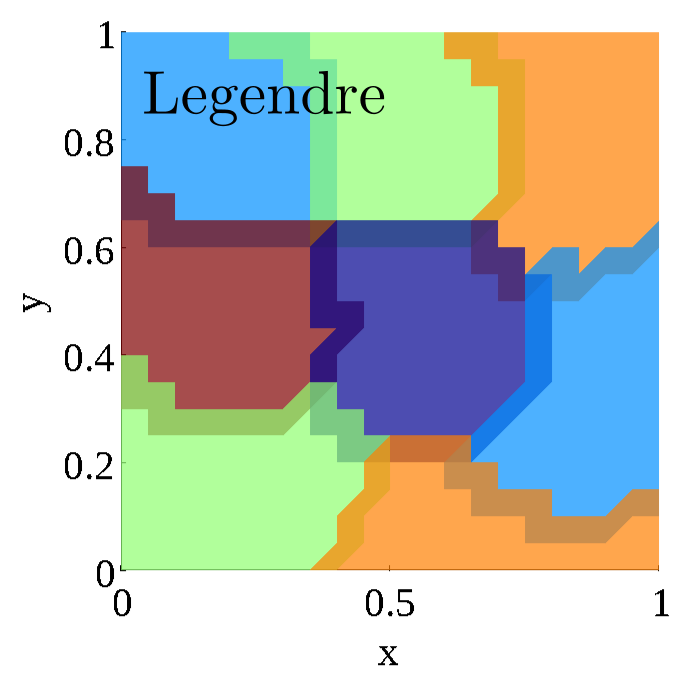}}%
					\subcaptionbox{ }[.45\linewidth]{\includegraphics[width=.55\linewidth]{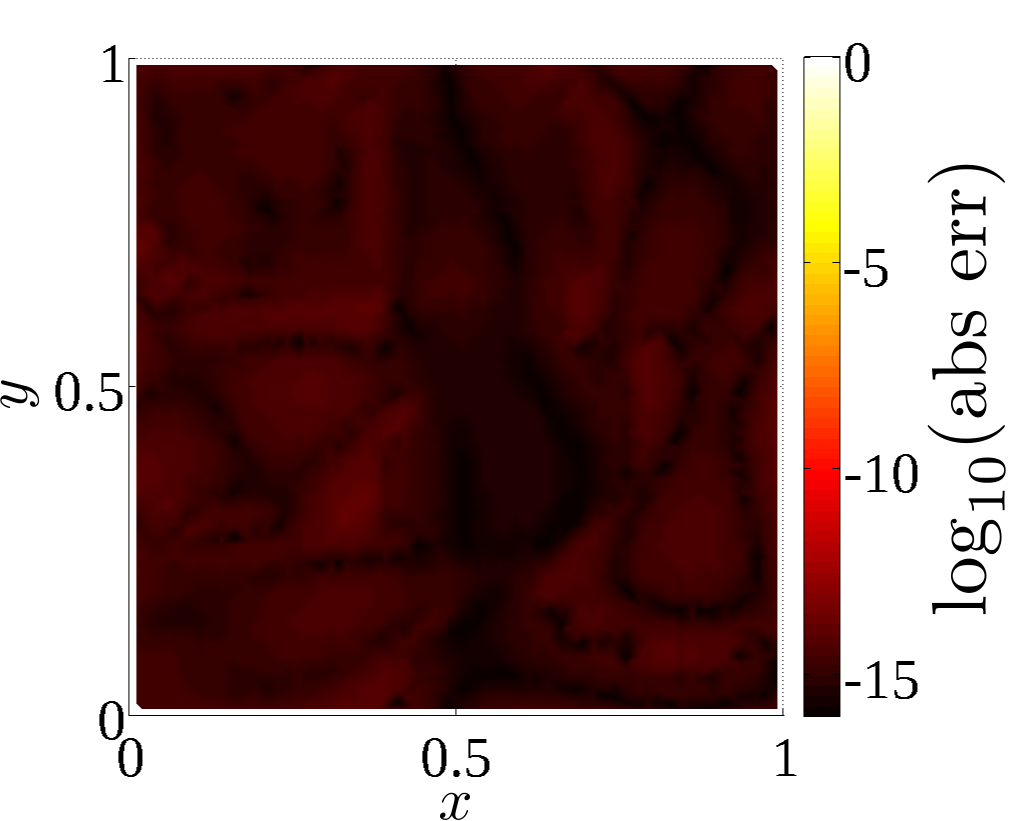}}
					\subcaptionbox{ }[.45\linewidth]{\includegraphics[width=.45\linewidth]{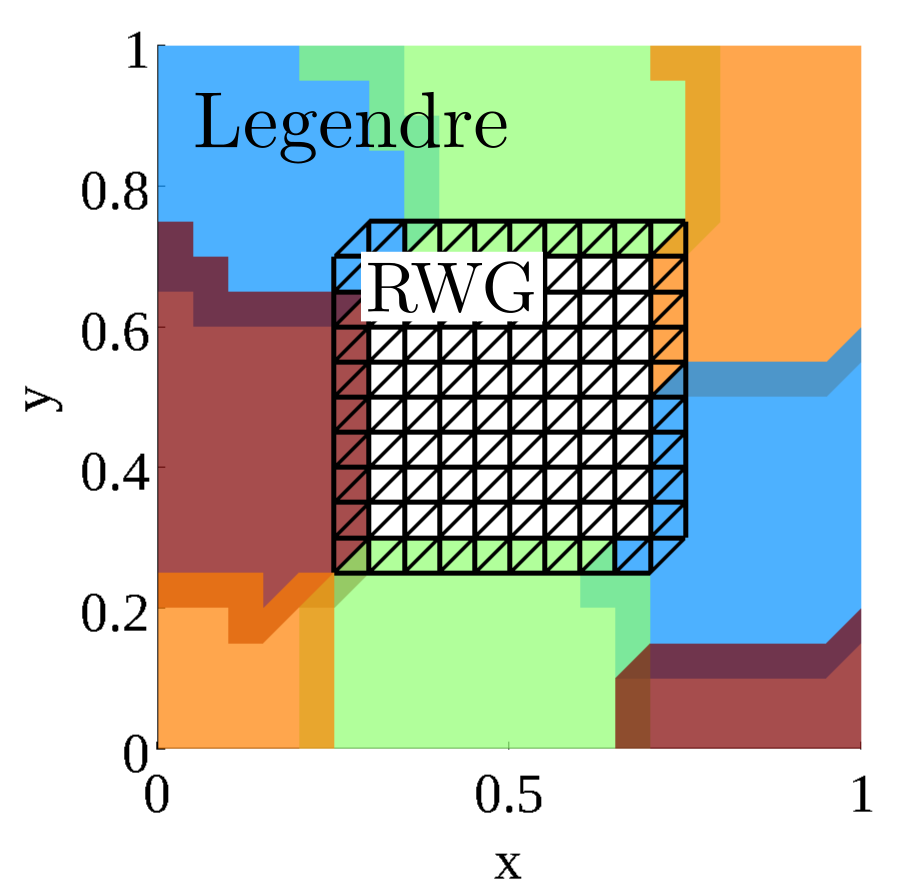}}%
					\subcaptionbox{ }[.45\linewidth]{\includegraphics[width=.55\linewidth]{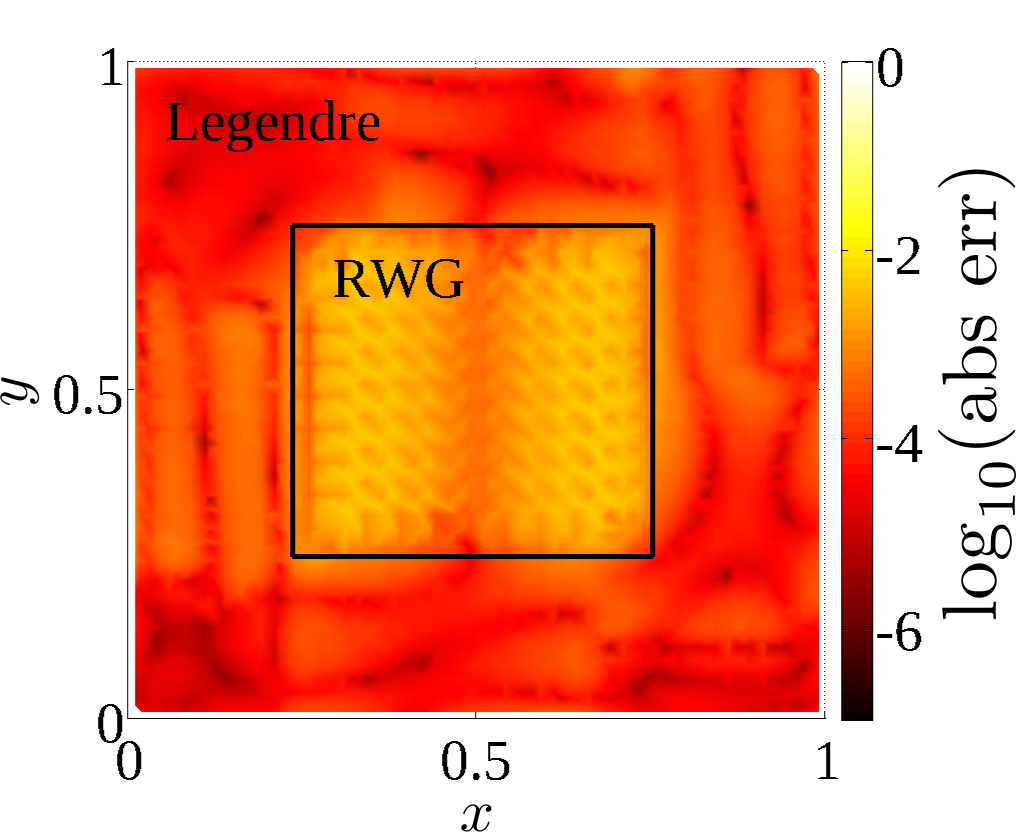}}
					
					\caption{Reconstruction error for a function $\mathbf{f}(x,y)=(x-.5)^2\uvec{x}$ using three different basis sets: all-RWG, all-Legendre, mixed RWG/Legendre.  (a)  All-RWG discretization, (b) All-RWG reconstruction error, (c) All-Legendre patch discretization (2nd order Legendre polynomials) (d) All-Legendre reconstruction error, (e) Mixed RWG/Legendre discretization  (2nd order Legendre Polynomials), (f) Mixed RWG/Legendre reconstruction error.}
					\label{fig:mixed_bas}
				\end{figure}

We begin with a representation (Gram) test in which a quadratic vector function $\mathbf{f}(x,y)=(x-.5)^2\uvec{x}$ is reconstructed on a flat $1$m$\times$$1$m plate using a mixture of a 2nd order Legendre polynomials defined on smooth patches and RWG functions defined on a triangular tessellation.  Figure \ref{fig:mixed_bas} shows the discretization and reconstruction error for three cases.  Reconstruction error is taken as the pointwise absolute error (abs err)$=|\tilde{\mathbf{f}}(x,y)-\mathbf{f}(x,y)|$, where $\tilde{\mathbf{f}}(x,y)$ is the reconstructed function.  

First, as a reference, we interpolate the analytical function $\mathbf{f}(x,y)$ using the standard linear RWG basis on a triangular tessellation with maximum edge length .07m.  This discretization is shown in figure {\ref{fig:mixed_bas}}.a. Figure {\ref{fig:mixed_bas}}.b shows the reconstruction error using the all-RWG basis set.  As one would expect, the error is highest at the edges where the RWG functions cannot reconstruct the normal component, and lowest where the argument of $\mathbf{f}(x,y)$ goes to zero (at $x=.5$m).  The purpose of this result is to provide a reference error level for RWG discretization for case 3, in which RWG and Legendre bases are mixed.
				
Figures {\ref{fig:mixed_bas}}.c and {\ref{fig:mixed_bas}}.d show discretization and reconstruction error for an all-Legendre polynomial basis of second order. Because the basis set is polynomial complete to the same order as $\mathbf{f}(x,y)$, the reconstruction is perfect to machine precision everywhere on the plate.  Most significantly, the reconstruction is perfect even in the overlap regions between patches where the partition of unity provides the continuity and blending between basis sets on neighboring patches.  Without the partition of unity, there would be significant error in the overlap regions.  

In figure \ref{fig:mixed_bas}.e, a $.25$m $\times$ $.25$m tessellated patch with RWG functions is placed in the center of the plate, while the outer boundary is discretized using flat patches supporting second order Legendre polynomials.  Figure \ref{fig:mixed_bas}.f shows the reconstruction error for the mixed RWG/Legendre basis set.  In this case, the error in the RWG patch matches that from the center of the plate in figure \ref{fig:mixed_bas}.b.  Outside of the RWG region, the error falls significantly, although due to the coupling between the RWG and Legendre patches, it does not decrease to the level of the all-Legendre solution.  Crucially, due to the blending effect of the partition of unity, no additional error is introduced in the transition region between the two basis sets.
\begin{figure}
	\begin{center}
		\subcaptionbox{\label{fig:sphLeg}}[.5\linewidth]{\includegraphics[width=.5\linewidth]{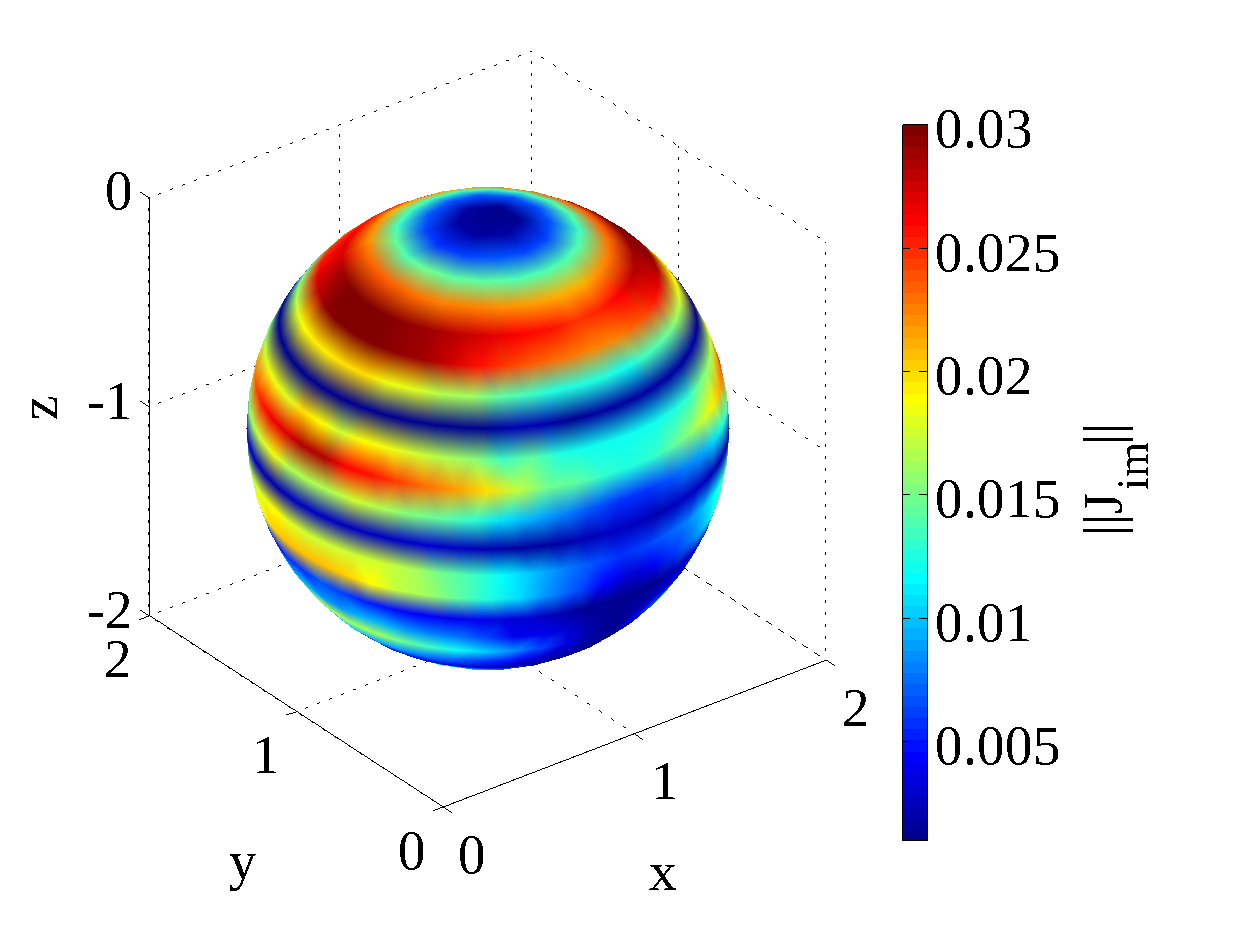}}%
		\subcaptionbox{\label{fig:sphRWG}}[.5\linewidth]{\includegraphics[width=.5\linewidth]{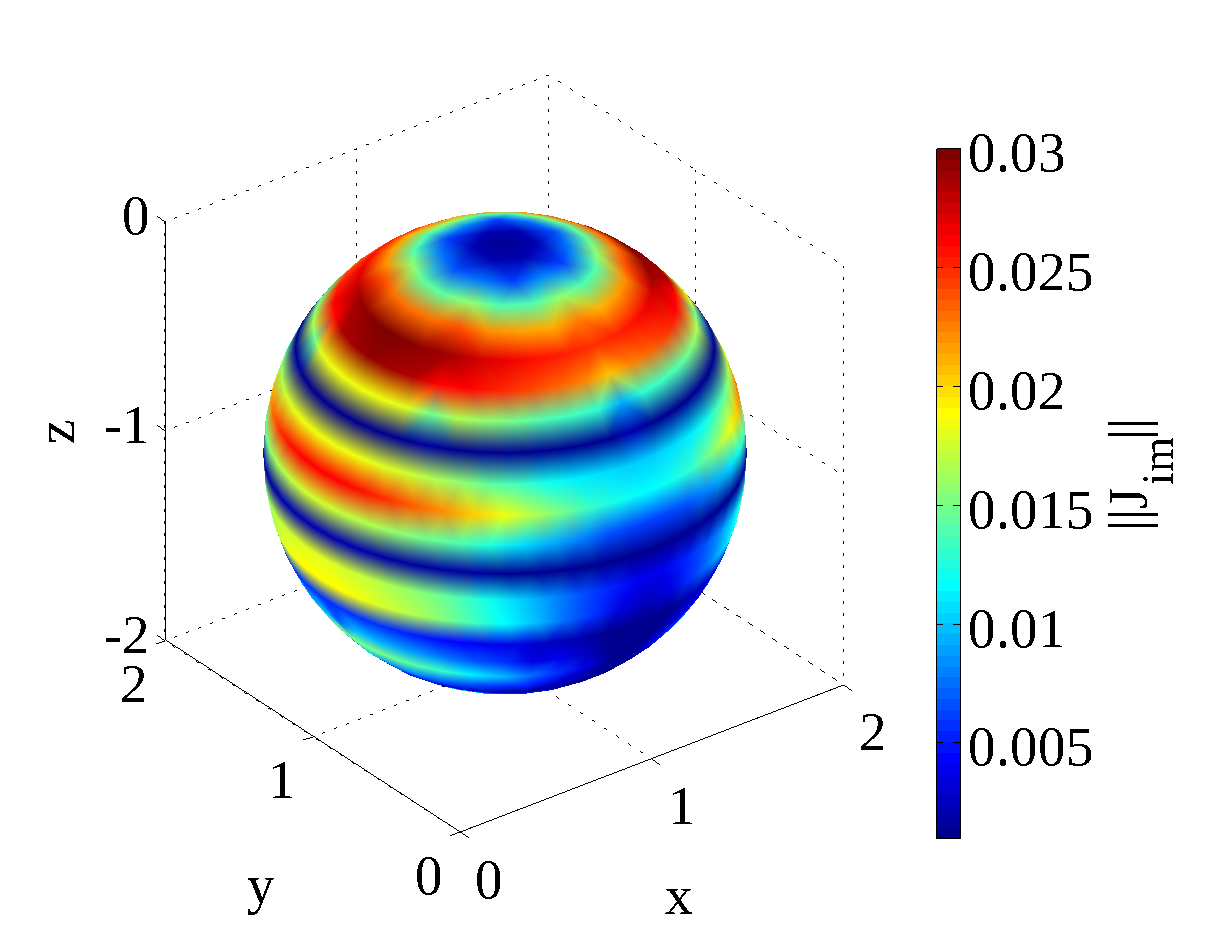}}
		\subcaptionbox{\label{fig:legPatch}}[.5\linewidth]{\includegraphics[width=.5\linewidth]{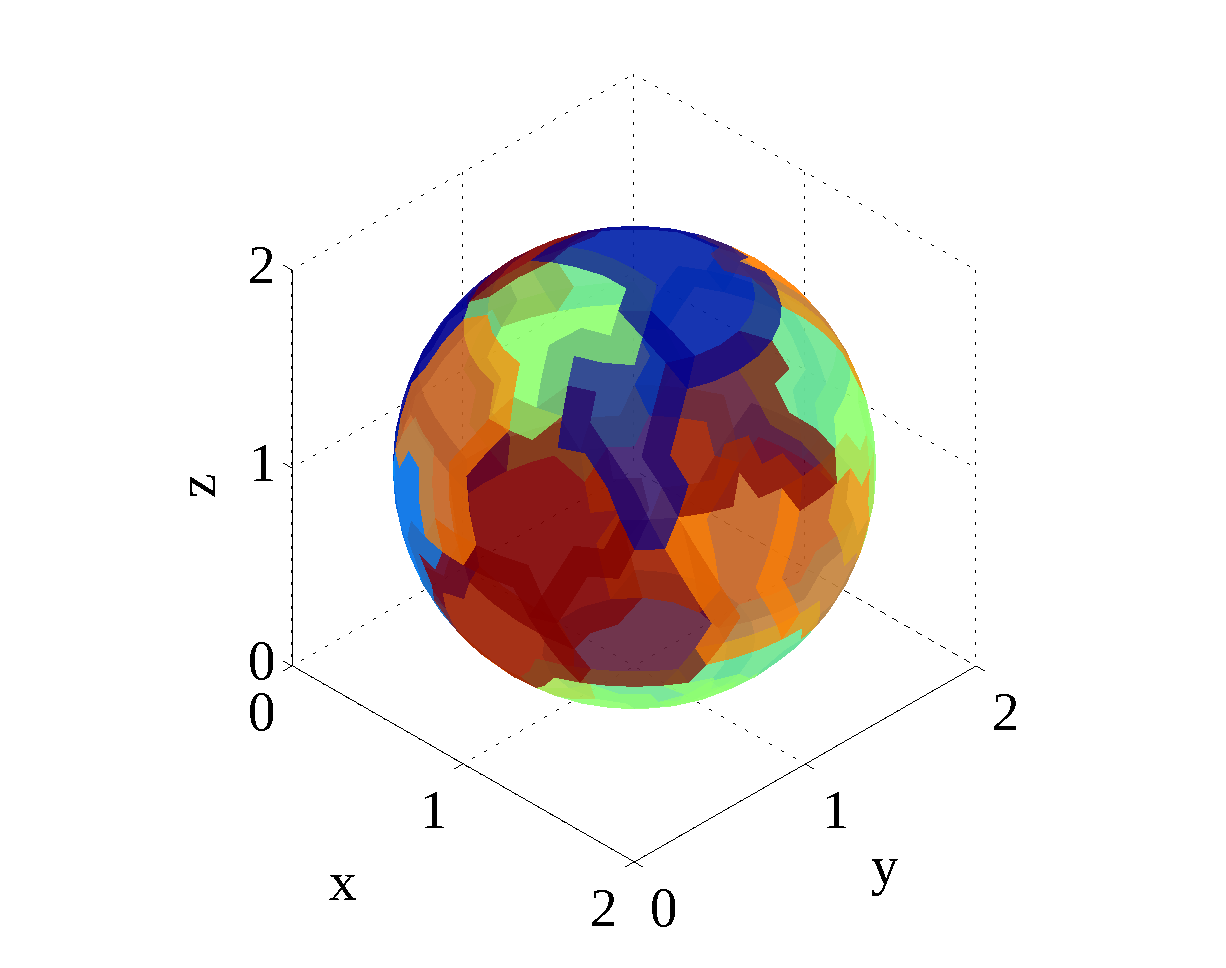}}
	\end{center}
	\caption{Surface currents induced by an incident plane wave on a sphere discretized with (a) overlapping GMM patches supporting a 4th order Legendre basis and (b) an RWG basis for reference.  Subfigure (c) shows the patches used to obtain the GMM result.  Magnitude of imaginary part of current $||J_{im}||$ is shown; $||J_{re}||$ yields similar plots.}
	\label{fig:sphCur}
\end{figure}

The next result shows currents induced on a spherical scatterer of radius $a=1\lambda$ due to an incident $x$-polarized plane wave propagating in the $-\uvec{z}$ direction.  Figure {\ref{fig:sphCur}}.a shows currents obtained via solving the MFIE using overlapping GMM patches supporting a 4th order Legendre basis, and figure {\ref{fig:sphCur}}.b shows the same result using RWG basis functions for reference.  Figure {\ref{fig:sphCur}}.c shows the overlapping patches used to obtain the GMM result.  The current densities show good agreement, although the higher-order GMM patches give smoother results relative to the RWG discretization.  Due to the analytical form of the scatterer, the sphere may be exactly represented using spherical local geometry descriptions in GMM.

\begin{figure}
		\begin{center}
		   \includegraphics[width=.5\textwidth]{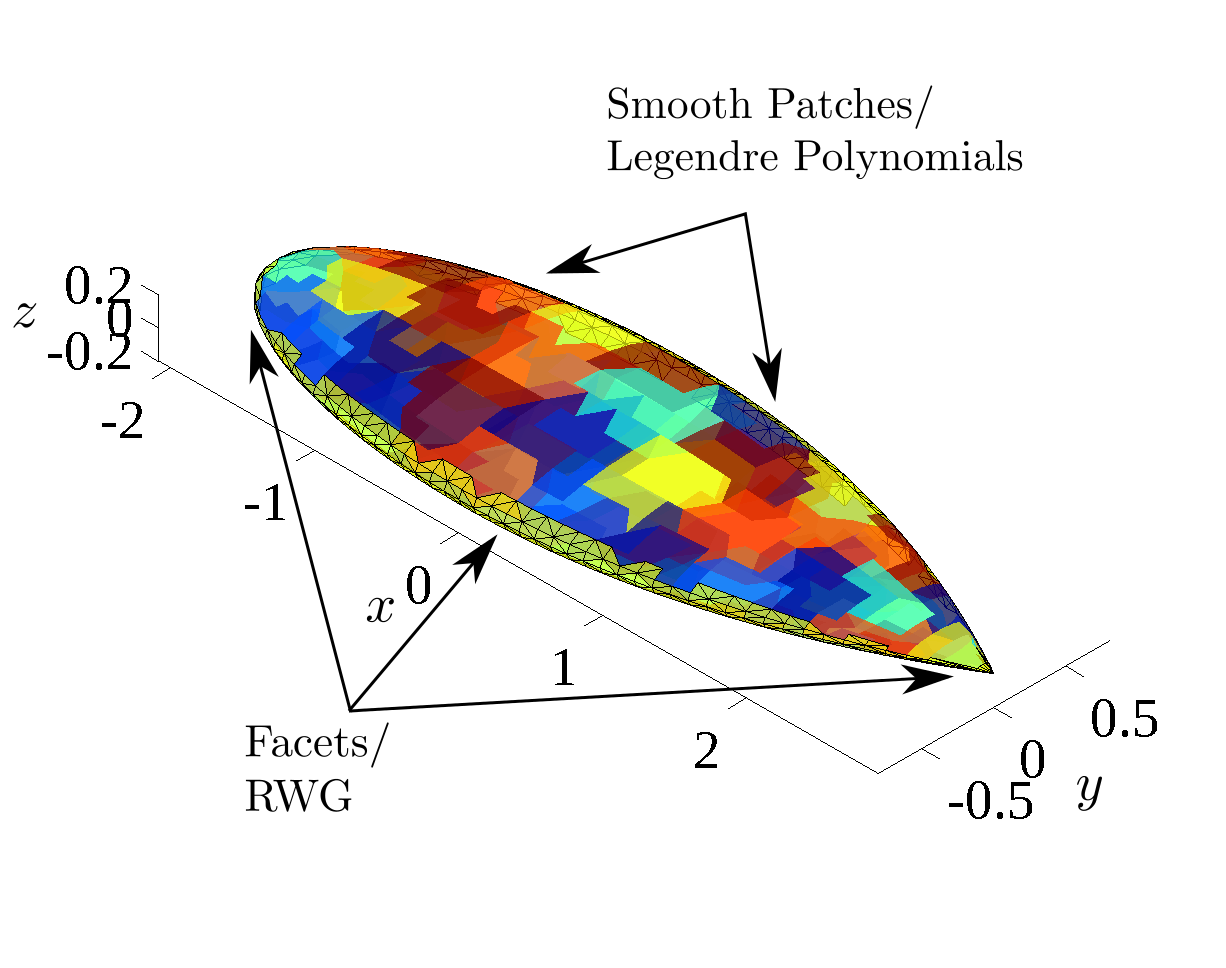}
			\caption{GMM discretization for a 4.5$\lambda$ NASA almond}
			\label{fig:almondpatches}
		\end{center}
	\end{figure}

\begin{figure}
		\begin{center}
		{\Large \scalebox{.5}{\fbox{\input{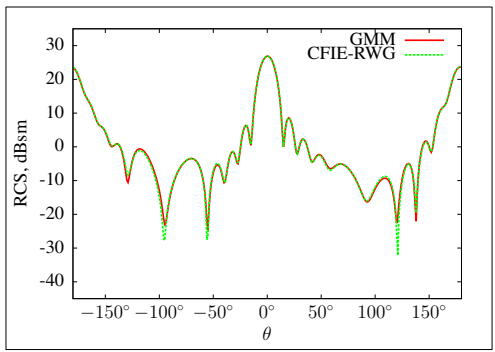}}}}
		\caption{NASA almond bistatic RCS, taken along the $\phi=0^{\circ}$ cut}
			\label{fig:almond1}
		\end{center}
	\end{figure}

The next scattering result demonstrates automated geometry and basis function assignment for a $4.5\lambda \times 1.5\lambda \times .4\lambda$ NASA almond, with the corresponding RCS compared against a well-validated CFIE-RWG reference solution.  The almond is illuminated by a unit amplitude $x$-polarized plane wave of frequency 300MHz incident in the $+z$ direction.  Here, a Legendre basis (eq. \ref{eq:lbas}) of maximum order $p=3$ is defined on 2nd order polynomial patches in the smooth regions, which are detected using Algorithm \ref{alg:bf}.  Figure {\ref{fig:almondpatches}} shows the discretization of the almond in terms of overlapping smooth and tessellated patches.  Tessellated patches supporting RWG basis functions are used to capture regions of high curvature and the geometric singularity at the tip.  One could envisage a tip basis function that does not rely on tessellation, but instead captures the tip singularity exactly.  While the design of such a basis is outside the scope of this paper, it could be seamlessly integrated into the GMM framework.  Figure \ref{fig:almond1} shows the bistatic RCS for the almond, which is taken along the $\phi=0^{\circ}$ cut.  The number of degrees of freedom for the CFIE-RWG solution is $N_{ref}=3636$ and the GMM result requires $N_{GMM}=1861$.

\begin{figure}[H]
			\begin{center}
			\includegraphics[width=.8\linewidth,trim= 0cm 0mm 0mm 1cm]{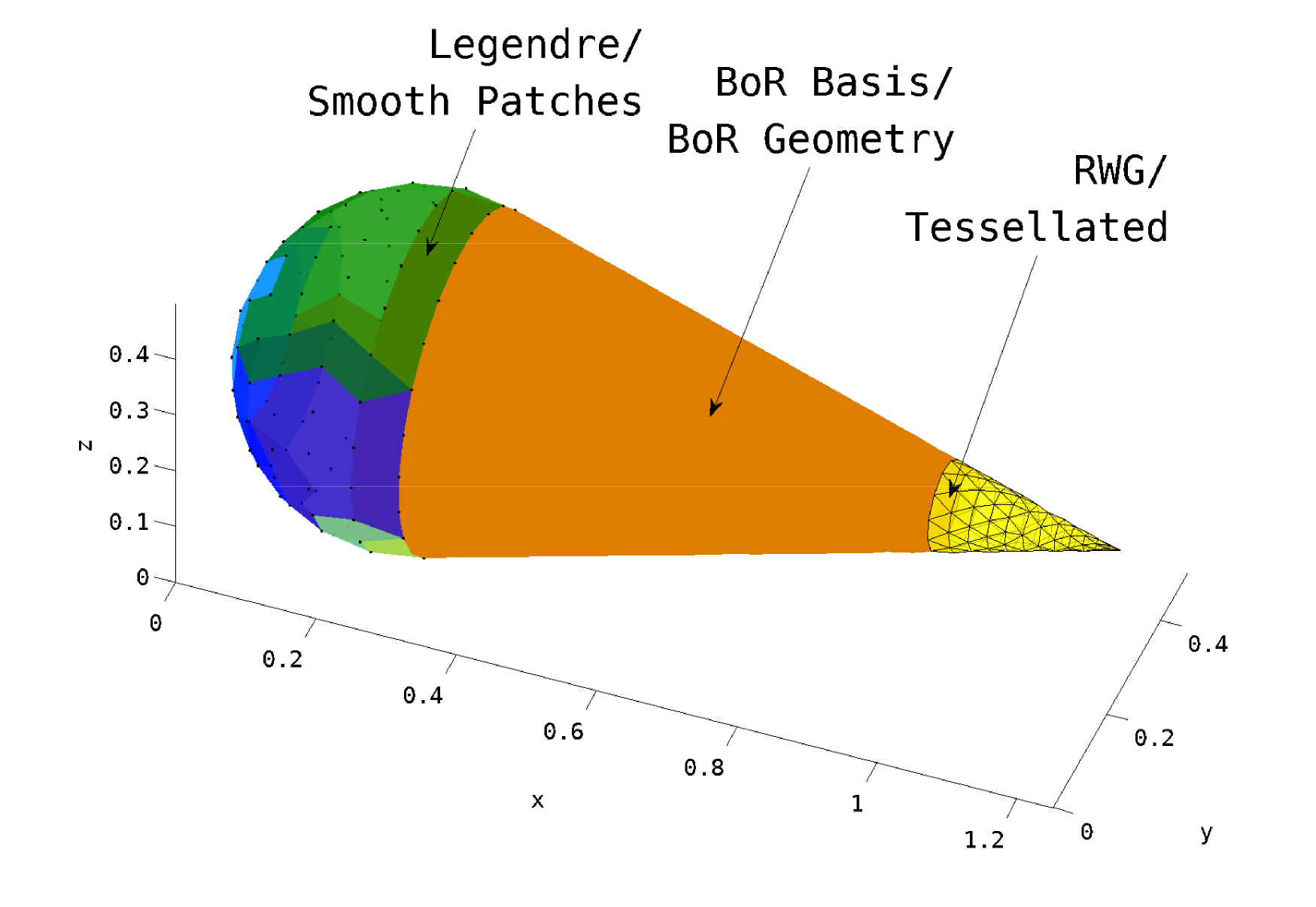}
			\caption{GMM discretization of a $1.2\lambda \times .4\lambda \times .4\lambda$ conesphere.\label{fig:cnsph2_geo}}
			\end{center}
	\end{figure}

\begin{figure}[H]
		{\center \Large \resizebox{.8\linewidth}{!}{\fbox{\input{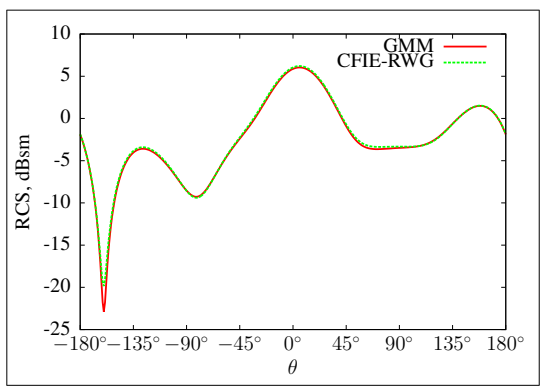}}}
		{\centering \caption{Bistatic RCS for conesphere at $300$MHz.  Wave is incident along the $\theta=0^{\circ}, \phi=0^{\circ}$ direction with $-x$-polarization.  RCS is taken along $\phi=0^{\circ}$ cut.\label{fig:cnsph2_rcs}}}}
	\end{figure}

	The next result, scattering from a $1.2\lambda \times .4\lambda \times .4\lambda$ diffraction-matched conesphere illuminated by a $-\uvec{x}$ polarized plane wave incident in the $\uvec{z}$ direction, illustrates the high degree of geometrical accuracy that may be obtained when the analytical form of the underlying scatterer is known.   Figure \ref{fig:cnsph2_geo} illustrates the GMM discretization for the conesphere.  The cone portion of the geometry (excluding the tip) is represented exactly through a single BoR patch which supports two types of BoR basis functions, defined as:
\begin{equation}
	\begin{split}
	\mb{f}_{m,1}(\rv)&=a P_n(\rho) e^{j m \phi} \rv_{\rho}(u_1,u_2) \\
	\mb{f}_{m,2}(\rv)&=a P_n(\rho) e^{j m \phi} \rv_{\phi}(u_1,u_2)
	\end{split}
\end{equation}
where $\rho(u_1,u_2)=\sqrt{u_1^2+u_2^2}$ and $\phi(u_1,u_2)=\arctan(u_2/u_1)$.  To capture the spherical portion of the scatterer to a high degree of precision, 5th order polynomial patches are employed with 3rd order Legendre basis functions.  The tip region is modeled with a tessellation and RWG functions.  Figure \ref{fig:cnsph2_rcs} shows a comparison of the GMM result against a reference CFIE-RWG result.  In this case, the GMM solution uses 126 Legendre functions on the spherical cap, 130 BoR basis functions on the barrel, and 294 RWG functions on the tip, so that the total GMM system size is $N_{GMM}=550$.  The reference RWG solution requires a total of $N_{ref}=1584$ degrees of freedom.  Agreement between the two results is quite good.


\begin{figure}[H]
			\begin{center}
			\begin{tikzpicture}
				\node[anchor=south west,inner sep=0] at (0,0) {\includegraphics[width=.8\linewidth,trim= 0cm 0mm 0mm 1cm]{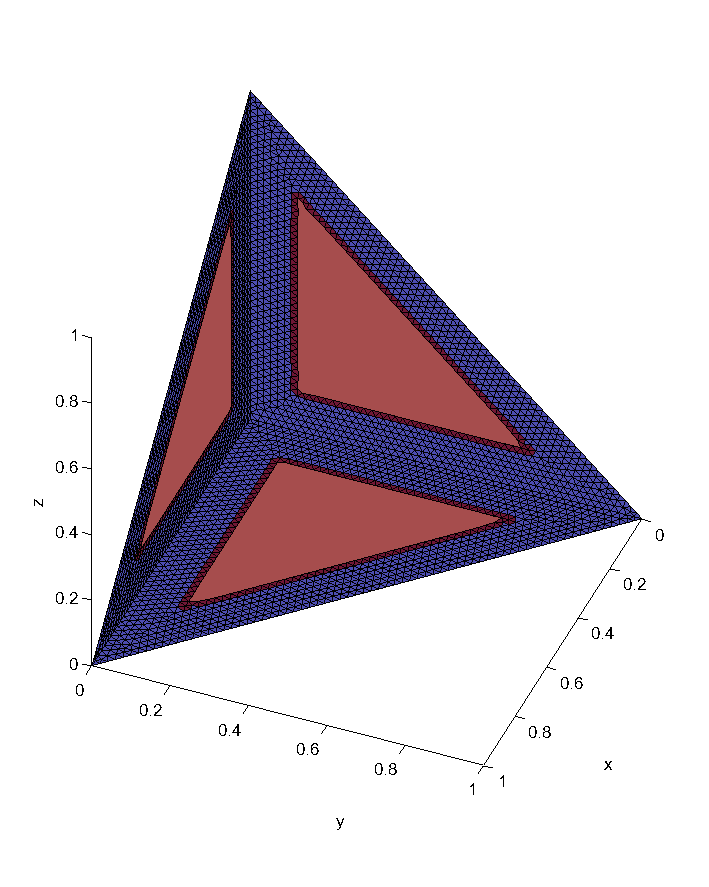}};
				\draw[thick,->] (4,2.3) node[anchor=north] {Plane Waves} -- (2,5);
				\draw[thick,->] (4,2.3) -- (4,5);
				\draw[thick,->] (4,2.3) -- (3.2,3.8);
				\draw[thick,->] (6,7) node[anchor=south west] {RWG} -- (6,4.2);
				\draw[thick,->] (6,7) -- (3,7.6);
			\end{tikzpicture}
			\caption{GMM discretization of a $5\lambda$ trihedral corner reflector\label{fig:tricor_geo}}
			\end{center}
	\end{figure}

	To illustrate incorporation of a PO-type basis, we show hybridization of RWG basis functions with plane wave functions in the discretization and solution of the $5\lambda\times5\lambda\times5\lambda$ trihedral corner reflector shown in figure \ref{fig:tricor_geo}.  The plane wave functions are given by:
\begin{equation}
	\begin{split}
		\mathbf{f}_{i,l,m}(\rv)=e^{j\mathbf{k}_{l,m}\cdot\rv(u_1,u_2)}\rv_{u_i},~~i=1,2
	\end{split}
\end{equation}
where $|\mathbf{k}_{l,m}|=\omega/c$ and $\uvec{k}_{l,m}=\sin\theta_l\cos\phi_m\uvec{e}_1+\sin\theta_l\sin\phi_m\uvec{e}_2+\cos\theta_l\uvec{e}_3$, with $\lbrace \uvec{e}_1,\uvec{e}_2,\uvec{e}_3 \rbrace$ a triad of orthogonal unit vectors defined with reference to the incident wave direction $\uvec{k}^i$ such that $\uvec{e}_3=-\uvec{k}^i$.  The solution requires only three plane wave patches on the interior of each side of the reflector, each of which is parameterized as a single plane using a $0$th order ``polynomial'' representation.  To accurately capture the variation in currents near geometric singularities and the corresponding diffraction lobes in the farfield pattern, an RWG border several elements wide is required.

\begin{figure}[H]
		{\center \Large \resizebox{.9\linewidth}{!}{\fbox{\input{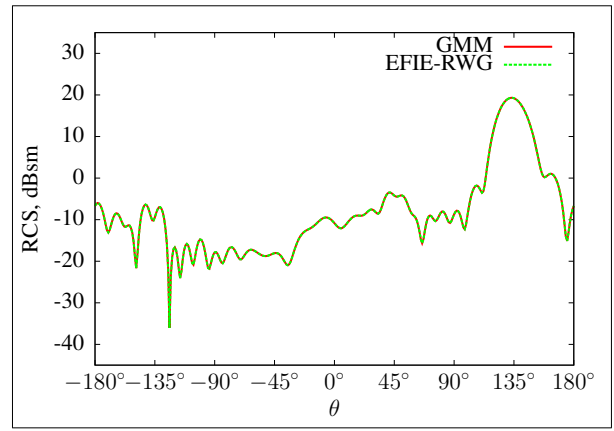}}}
		{\centering \caption{Bistatic RCS for trihedral corner reflector at $300$MHz.  Incident wave is traveling in $\theta^i=45^{\circ}, \phi^i=45^{\circ}$ direction, with $x$ polarization relative to incidence direction.  RCS is taken along the $\phi=45^{\circ}$ cut}\label{fig:tricor_rcs}}}
	\end{figure}

The bistatic RCS shown in figure \ref{fig:tricor_rcs} was obtained using a total of 144 plane wave bases (48/patch) and 8602 RWG bases, for a total of  $N_{GMM}=8746$ degrees of freedom.  The reference RWG system used $N_{ref}=12918$ degrees of freedom.  The GMM results in this case compare particularly well with the reference code because the reference tessellation matches the GMM geometry representation exactly.   The RCS result in figure \ref{fig:tricor_rcs} is taken along the $\phi=45^{\circ}$ cut.

\begin{figure}	
						\subcaptionbox{}[.5\linewidth]{\includegraphics[width=.5\linewidth]{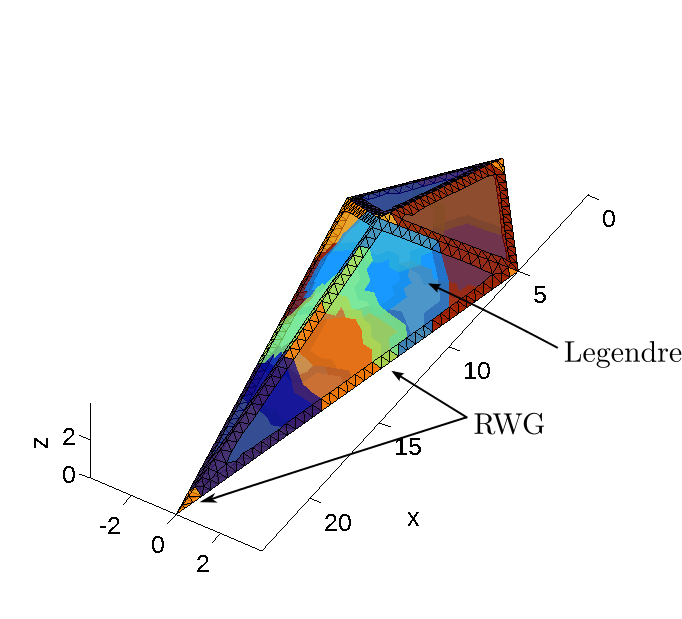}}%
						\subcaptionbox{}[.5\linewidth]{\includegraphics[width=.5\linewidth]{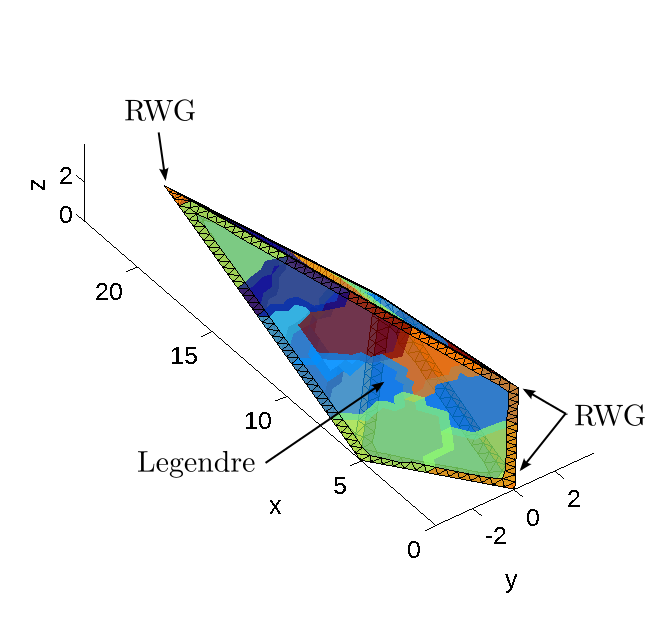}}
						\caption{GMM discretization of a $5\lambda\times1.5\lambda\times.8\lambda$ arrow (a) top view and (b) bottom view}
						\label{fig:arr_geo}
					\end{figure}

\begin{figure}
						{\center \large \resizebox{.8\linewidth}{!}{\fbox{\input{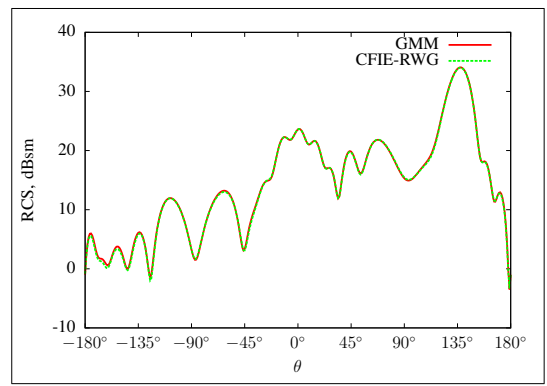}}}
						\caption{Arrow bistatic RCS at $64$MHz. Wave is incident in the $\theta_i=135^{\circ}, \phi_i=20^{\circ}$ direction and is $x$-polarized relative to incidence direction.  RCS is taken along the $\phi=0^{\circ}$ cut.}
						\label{fig:arr_rcs}}
					\end{figure}

		To demonstrate the reduction in system size that results from using a polynomial bases on a faceted scatterer possessing large flat regions, we present a scattering result from an arrow geometry that possesses several sharp features in addition to large flat faces.  The scatter measures $5\lambda\times1.5\lambda\times.8\lambda$ at $64$MHz.  In this case the scattering body is discretized with a combination of large flat patches supporting a $4$th order Legendre basis on the flat sides and triangular tessellation/RWG on the corners and tips.  Figure \ref{fig:arr_geo} shows the resulting patches and tessellation.  The bistatic RCS in Figure \ref{fig:arr_rcs} due to an $x$-polarized plane wave incident along the $\theta_i=135^{\circ}, \phi_i=20^{\circ}$ direction was obtained with the CFIE, $\alpha=.5$, and closely matches the reference solution.  The number of unknowns is reduced from $N_{ref}=3549$ for the CFIE-RWG solution to $N_s=1830$ using the GMM mixed Legendre/RWG solution.

\begin{figure}
				\subcaptionbox{}[1.0\linewidth]{
					\includegraphics[width=.7\linewidth]{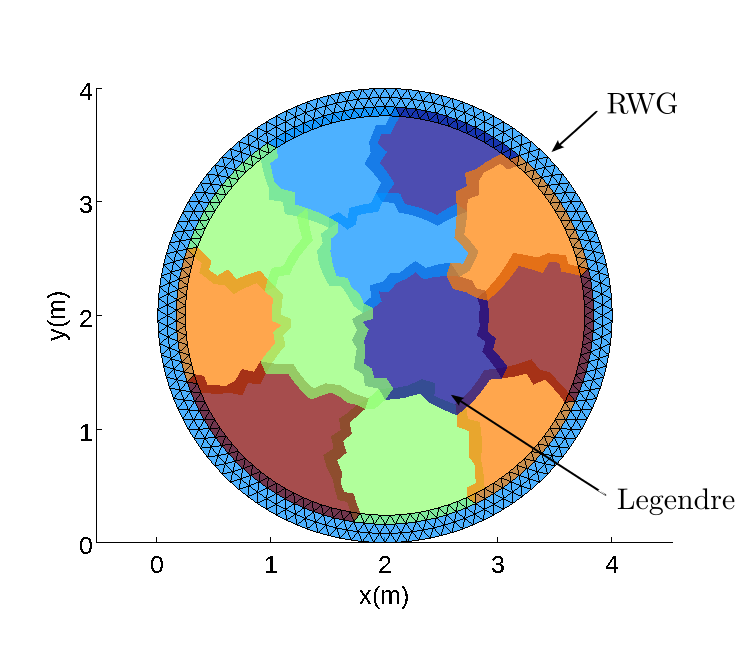}};
				\subcaptionbox{}[1.0\linewidth]{
				\includegraphics[width=.7\linewidth]{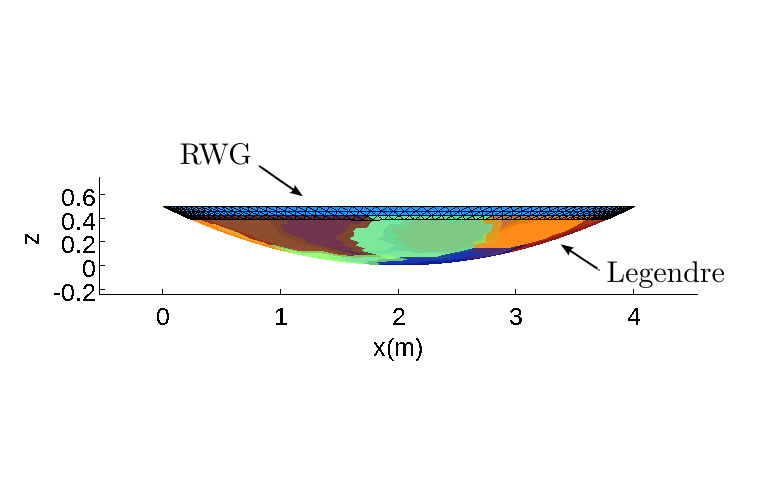}
					}
					\caption{GMM discretization for a $4\lambda\times.35\lambda$ Parabolic Reflector (a) top view and (b) side view}
					\label{fig:par_geo}
		\end{figure}

\begin{figure}
			\center \large \resizebox{.9\linewidth}{!}{\fbox{\input{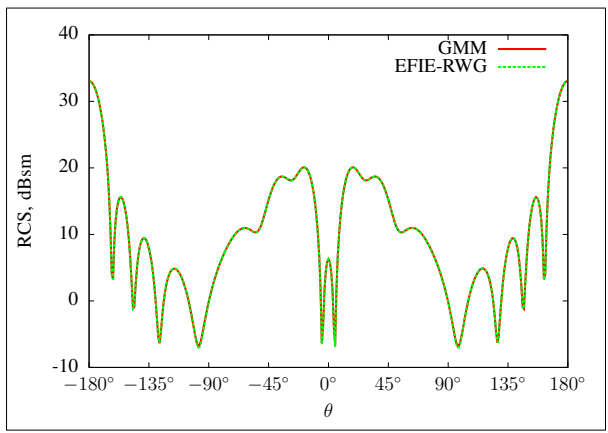}}}
			\caption{Parabolic reflector bistatic RCS. Wave is incident in the $\theta_i=180^{\circ}, \phi_i=0^{\circ}$ direction with $x$-polarization.  RCS is taken along the $\phi=0^{\circ}$ cut.}
			\label{fig:parrcs}
		\end{figure}

					Finally, we analyze the parabolic reflector in Figure \ref{fig:par_geo}, which is an open curved structure.  The smoothly curved interior is modeled using 4th order smooth polynomial patches, and the rim is captured by a piecewise flat triangular tessellation.  A 5th order Legendre basis is used on the polynomial patches, and an RWG basis is employed on the tessellated portion.  The RCS result in figure \ref{fig:parrcs} shows scattering due to an $x$-polarized plane wave with incidence angles $\theta_i=180^{\circ}, \phi_i=0^{\circ}$.  Again, the agreement in RCS between the reference EFIE-RWG and GMM solutions is excellent.  For the reflector, the EFIE-RWG code required $N_{ref}=4494$ degrees of freedom, and the GMM solution required $N_{GMM}=1638$ degrees of freedom.

\section{Conclusion}
\label{s:Con}

This work extends the scope of the Generalized Method of Moments to include multiple geometry representations and mixed current approximation spaces on PEC scatterers. Decomposition of the scatter surface via overlapping patches and the partition of unity allow local geometry descriptions capable of handling geometrical features including smooth regions, regions for which a priori functional description is known or can be easily extracted, and regions including geometrical singularities.  Furthermore, the partition of unity permits mixtures of multiple classes of Entire Patch and/or Sub Patch basis sets within a single simulation.  In particular, the introduction of subpatch basis sets on traditional tessellations allows straightforward handling of geometrical singularities, even in problems with smooth higher order geometry descriptions.  Finally, the entire process of geometry and basis assingment can be automated for complex structures.  The resulting method permits discretization the underlying integral operators in a manner that more closely matches the physics and may result in significant reductions in the number of degrees of freedom required for a given problem relative to traditional moment method solvers.

Although the computational costs associated with assembling and solving a GMM system are highly dependent on the particular mixture of basis sets and patch sizes employed, the cost of the algorithm in both complexity and storage is similar to that of extant Moment Methods.  If currents are predominantly discretized with entire patch basis functions, the cost scales as that of a mapped higher-order Moment Method; alternatively, if sub-patch basis functions are the principle basis type, the cost approaches that of traditional tessellation-based schemes such as RWG, rooftop, or GWP bases.  The only added cost of evaluating matrix elements in GMM relative to traditional Moment Method schemes is that of computing the partition of unity, and since the partition of unity on each patch is non-unity only in the overlaps between patches, its evaluation adds very little to the overall complexity.

Future investigations will expand the types of basis functions and local geometry parameterizations used in GMM and apply GMM to dielectric problems.  One interesting possibility is to investigate GMM in the context of a domain decomposition method (DMM).   GMM is a discretization method that stitches together different basis and geometry descriptions to form a Moment Method system, whereas DDM is a solution method in which the Moment Method system is solved by breaking the overall problem into subproblems and solving each smaller problem individually, subject to global consistency constraints.  The two methods therefore address different aspects of the discretization and solution of electromagnetic integral equations, and could possibly complement each other well.  Using a GMM discretization in a domain decomposition framework would yield a method in which individual subdomains could be solved independently as in {\cite{Peng2011} or the Equivalence Principle Algorithm in \cite{Li2006c}, but where continuous transitions between non-conformal subdomains are provided by the partition of unity.

\subsection*{Acknowledgements}
The authors wish to acknowledge the HPCC facility at Michigan State University and support from NSF CCF-1018516 and NSF CMMI-1250261.  D. Dault would like to acknowledge support from the NSF Graduate Research Fellowship Program.


\end{document}